\documentclass[12pt]{article}
 
\usepackage{amssymb}
\usepackage{amsmath}
\usepackage{amsfonts}
\usepackage{amscd}
\usepackage{graphicx}
\usepackage{amsthm}
\usepackage{setspace}
\usepackage{epsfig} 
\usepackage{bm}
\usepackage{amssymb}
\usepackage[authoryear,round]{natbib}
\usepackage{multirow}
\usepackage{rotating}

\newtheorem{theorem}{{\bf \sc Theorem}}
\newtheorem{proposition}{{\bf \sc Proposition}}
\newtheorem{fact}{{\bf \sc Fact}}
\newtheorem{lemma}{{\bf \sc Lemma}}
\newtheorem{corollary}{{\bf \sc Corollary}}

\newtheorem{definition}{{\bf \sc Definition}}

\newcommand{\be}{\begin{equation}}
\newcommand{\ee}{\end{equation}}
\newcommand{\bes}{\begin{equation*}}
\newcommand{\ees}{\end{equation*}}

\def\eproof{\hfill \hbox{\hskip3pt\vrule width4pt height8pt depth1.5pt}}
\newcommand{\R}{\mathbb R}

\newcommand{\Ex}{\mathbb E}
\renewcommand{\Pr}{\mathbb P}

\DeclareMathOperator{\abs}{abs}

\DeclareMathOperator{\DWH}{DWH}
\DeclareMathOperator{\EDWH}{EDWH}
\DeclareMathOperator{\MT}{MT}
\DeclareMathOperator{\CT}{CT}

\allowdisplaybreaks

\begin{document}

\title{\textbf{How Homophily Affects \\ Learning and Diffusion in Networks}\thanks{%
Jackson gratefully acknowledges financial support from the
NSF under grant SES--0647867. Golub gratefully acknowledges financial support from an NSF Graduate Research Fellowship, as well as the Clum, Holt, and Jaedicke fellowships at the Stanford Graduate School of Business. We thank James Moody for making available the network data
from the `Add Health' data set. We also thank Christoph Kuzmics, Carlos Lever, and Amin Saberi for comments on presentations.}}
\author{Benjamin Golub\thanks{\mbox{Graduate School of Business, Stanford University.
Email: bgolub@stanford.edu,}\hfill \mbox{http://www.stanford.edu/$\sim$bgolub/}.} \and %
 \and Matthew
O. Jackson\thanks{%
Department of Economics, Stanford University and external faculty of the Santa Fe Institute.
Email: jacksonm@stanford.edu, http://www.stanford.edu/$\sim$jacksonm/} }
\date{First Arxiv Version arXiv:0811.4013v1: November 25, 2008 \\
Expanded Version: February 7, 2009}
\maketitle

\begin{abstract}

We examine how three different communication processes operating through social networks are affected by homophily -- the tendency of individuals to associate with others similar to themselves. Homophily has no effect if messages are broadcast or sent via shortest paths; only connection density matters. In contrast, homophily substantially slows learning based on repeated averaging of neighbors' information and Markovian diffusion processes such as the Google random surfer model. Indeed, the latter processes are strongly affected by homophily but completely independent of connection density, provided this density exceeds a low threshold. We obtain these results by establishing new results on the spectra of large random graphs and relating the spectra to homophily. We conclude by checking the theoretical predictions using observed high school friendship networks from the Adolescent Health dataset.

\smallskip

Keywords: networks, learning, diffusion, homophily, friendships, social networks, random graphs,
mixing time, convergence, speed of learning, speed of convergence

JEL Classification Numbers: D83, D85, I21, J15, Z13
\end{abstract}

\newpage

\section{Introduction}

How does a society's structure affect the speed at which information diffuses within it? In particular, how do segregation patterns in a social network affect how information is diffused and aggregated within that society?   How does that relationship change as we vary the communication process? In this paper, we take a step toward answering these questions by studying how information transmission is affected by homophily, an almost universally observed feature of social networks.

Homophily -- the tendency of individuals to associate with those similar to themselves -- has been observed since antiquity\footnote{By Plato's \nocite{Plato} time, homophily was already considered proverbial. In \emph{The Republic}, Cephalus says: ``For it often happens that some of us elders of about the same age come together and verify the old saw of like to like'' (Book I, p. 329).} and studied intensively by sociologists under that name since Lazarsfeld and Merton \citeyearpar{Lazarsfeld-Merton}. It has been documented across a wide array of different characteristics, including race, age, ethnicity, profession, religion, and various behaviors.  Indeed, homophily is one of the most pervasive and robust tendencies of social networks (see McPherson, Smith-Lovin and Cook \citeyearpar{homophily-survey} for a survey).

Learning and diffusion processes in networks have also been the focus of many recent studies in economics and related fields.\footnote{See, for instance, Ellison and Fudenberg \citeyearpar{EllisonFudenberg2}, Bala and Goyal \citeyearpar{BalaGoyal1},  DeMarzo,  Vayanos, and Zwiebel \citeyearpar{PersuasionBias}, Gale and Kariv \citeyearpar{GaleKariv},
Banerjee and Fudenberg \citeyearpar{WordOfMouth}, Golub and Jackson \citeyearpar{GolubJackson2007}, Acemoglu, Dahleh, Lobel and Ozdaglar \citeyearpar{AcemogluRational}, and Acemoglu, Nedic, and Ozdaglar \citeyearpar{AcemogluRuleOfThumb}.} These derive results about the convergence, speed, and/or accuracy of various communication processes, which include the diffusion of information, the formation of consensus, and other forms of learning and communication that are integral to social and economic behaviors. 

Despite the activity in these two related fields, there has been effectively no modeling of the impact of homophily on learning or diffusion processes. In this paper, we address this gap by modeling both homophily and communication explicitly and using the model to examine how homophily affects communication in various settings. 

% In doing this, we bring together two important but previously separate threads of research: statistical models of networks with homophily and models of information diffusion and learning in a given network. Random network modeling allows us to capture homophily in a natural way: certain links are more likely to form than others but the precise realization of the network is random.  Working with models of communication and learning in social networks allows us to understand how these are affected by changes in homophily.

We use a probabilistic model of homophily that we call the \emph{multi-type random network}. It generalizes the seminal Erd\H{o}s-R\'{e}nyi random network model and nests several other models as well. In our model,  agents are divided into different types, and then links are formed independently between various agents, with the probability of a link forming between two agents depending on the types of the agents involved.\footnote{The probabilities governing the linking may arise in various ways -- through a process of choices, or through differential opportunities for meeting various types, or some combination; this is modeled explicitly in Currarini, Jackson, and Pin \citeyearpar{CurrariniJacksonPin}. Here we abstract away from these issues and take the structure of linking probabilities as given.}
Once the network is generated, it remains fixed as learning or diffusion occurs. We consider three different processes. Whether or not the homophily in the network structure affects the speed of the learning or diffusion turns out to depend on the type of learning or diffusion process.

The first process we study is one in which information is either broadcast or navigated to its destination via shortest paths. This class includes many peer-to-peer systems like the Internet, mechanisms where messages are routed within an organization using something like an organizational chart, and information spreading phenomena where people tell an important piece of news to everyone they know. The second is a process based on linear updating or learning, first modeled by French \citeyearpar{French} and Harary \citeyearpar{Harary}, where individuals update their  beliefs or actions by repeatedly taking weighted averages of their neighbors' beliefs or actions.  This captures boundedly rational processes of updating as well as pressures to conform or a desire to match actions of neighbors.\footnote{Such updating processes can lead a society to an optimal aggregation of information in some settings, depending on the specifics of the social network structure (e.g., see Golub and Jackson \citeyearpar{GolubJackson2007}).}
The third is a random walk process on a network, where some particle hops around the network, having equal probability of moving along any link out of its current node; one example is Google's famous model of a surfer who randomly follows a link out of the website he is currently visiting.
These three processes encompass many important forms of network-based communication and diffusion.

Our main results show that whether or not homophily has an impact depends on the communication process, and we detail precisely how homophily matters when it does. In particular, we show that processes based on shortest paths are unaffected by homophily, while averaging processes and random walks are affected and can be substantially slowed down by it. The reason that homophily does not affect average shortest path lengths is that even with substantial homophily and the resulting clustering, the number of vertices that can be reached in $t$ steps is exponential in $t$. Homophily may change who is close and who is far from a given agent, but does not change the expected distance between two random nodes in the network. In contrast, processes based on weighted averaging or random walks are substantially slowed down as homophily increases: even though the average path length is unchanged, there are relatively fewer paths between agents of different types as homophily increases. This means that a node is more influenced by others of its own type, which reinforces global heterogeneity in beliefs or behaviors and slows down convergence to a steady state. 

In contrast, increasing link density speeds up shortest path communication, but has no effect on the speed of linear updating processes once density exceeds a low threshold. The first point is clear: adding links can only reduce distances, and will often do so dramatically by creating ``shortcuts'' which allow fast traversal of the network. To see why increasing link density in a homophilous network does not reduce delay in reaching consensus, consider a network in which a typical agent has nine friends of her own type and one friend of a different type. Suppose such an agent is in an island that disagrees with the rest of the network. At every step, she is pulled toward the global consensus somewhat by her outside friend ($10\%$ of the influence on her beliefs) but pulled strongly back by her own island. Thus, consensus is slow.  Now we double the number of links but hold the homophily fixed. So the typical agent will now have eighteen friends of her own type and two of a different type. As before, only $10\%$ of her beliefs will be coming from outside, so she will be pulled toward consensus equally slowly. 

\begin{table}[htbp]\centering
 \caption{A qualitative summary of our main results on how communication speeds in the processes we study are affected by homophily and link density.}
\label{tab:bigpicture}
\vspace{.2in}
\begin{tabular}{c c c c }
 & & \multicolumn{2}{c}{\textbf{\emph{Independent Variable}}} \\ 
  &  &   Density &   Homophily \\ 
\multirow{3}{16mm}{\hfill \begin{sideways}\parbox{16mm}{\textbf{\emph{Process}}}\end{sideways}}&  \rule[-6mm]{0cm}{12mm} Shortest Path   &  $\uparrow$ & $0$ \\
 &  Linear Updating  & \multirow{2}{*}{\rule[-6mm]{0cm}{12mm} $0$} & \multirow{2}{*}{\rule[-6mm]{0cm}{12mm} $\downarrow$} \\
 &  and Random Walk & & \\
\end{tabular}\end{table}

A qualitative summary of these relationships appears in Table \ref{tab:bigpicture}. An interesting implication is the following.   Consider a model where agents form links to others through a random search process, such as the one discussed in Currarini, Jackson and Pin \citeyearpar{CurrariniJacksonPin}. Suppose that we consider a change in the matching technology -- such as the introduction of social networking software -- so that it becomes easier to search for agents of one's own type. If agents have some preference for connecting to agents of their own types, this would lead to an increase in the overall density of links in the network, but would also cause the typical agent to spend a larger fraction of links on agents of the same type.  What would the ultimate impact be in terms of communication?  Our results imply that shortest-path based communication would become faster, but at the same time, Markovian processes of communication such as the linear updating process would converge more slowly!

As an empirical illustration of the results, we examine the time to convergence of these processes on social networks from over eighty different high school friendship networks that exhibit varying degrees of homophily.   We show that the predictions of our analysis fit well and that the speed of convergence depends on homophily in the predicted ways.

\subsection*{Conceptual Outline}

Our results involve several layers, including some contributions on the mathematical side which are needed to deduce the relationships discussed above.  In particular, in order to relate network structure (including homophily) to convergence speeds we have to work with the spectral decomposition of a network, which naturally leads us to develop new results on the spectra of large random graphs.  Given this layering, it is useful to have a road map of how all of our results fit together. 

After introducing the model and background definitions, we first present the main conceptual results that relate homophily to speed of communication.  
Having the main conclusions in hand, we then present a series of results that are used to derive those conclusions.   In particular, we begin with a statement of standard results relating 
speed of convergence to second eigenvalues.   Next, we present our key technical theorem, showing that the second eigenvalue associated with a random network will be close to the second eigenvalue of a smaller matrix which deals only with relative linking probabilities across types.   That is, all that really matters in determining the second eigenvalue in large societies is the expected connection probabilities between types.  This result allows us to  derive second eignvalues for the multi-type random networks based on homophily patterns and relate the eigenvalues to simple measures of homophily.  Thus, the key technical result allows us to tie homophily to second eigenvalues, which in turn govern consensus and mixing times.  
Putting all of this together provides our conceptual conclusions.

\section{The Model: Networks and Processes}

\subsection{Networks}
 \label{sec:networks} 
Given a set of $n$ nodes $N= \{1,\ldots,n\}$, a network is represented via its adjacency matrix: a symmetric $n$-by-$n$matrix $\mathbf{A}$ with entries in $\{0,1\}$. 
The interpretation is that $A_{ij} = A_{ji} = 1$ indicates that nodes $i$ and $j$ are linked, and we restrict attention to undirected networks.\footnote{Although we conjecture that the results can be extended to directed networks without much change in the statements (as the communication/learning processes have direct extensions to the directed case), there are parts of the proofs that take advantage of the symmetry of the adjacency matrix, and so we are not sure of what modifications would ensue in examining directed networks.} 

Let $d_i(\mathbf{A}) = \sum_{j=1}^n A_{ij}$ denote the degree of node $i$. Let $d_{\min}(\mathbf{A})$ and $d_{\max}(\mathbf{A})$ be the minimum and maximum degrees, respectively, and $\bar{d}(\mathbf{A})$ denote average degree, and let
$$D(\mathbf{A})=\sum_i d_i(\mathbf{A})$$
be the total degree in the society.

\subsection{Multi-Type Random Networks} 
\label{sec:multitype}

In order to study the impact of homophily on communication and learning through a network, we introduce a random network model that incorporates homophily. 
The seminal random network model of Erd\H{o}s-R\'{e}nyi random networks is a special case of the model here (and the same is true of the model based on degree distributions of Chung and Lu \citeyearpar{chunglu}) which allows us to make benchmark comparisons to the literature on speeds of processes on networks without homophily.  

The structure we use to model homophily is what we call the \emph{multi-type random network}.\footnote{This can be seen as a variant on statistical models that have been used to capture homophily in networks, such as various $p^*$ models (e.g., see the references and discussion in Jackson \citeyearpar{JacksonBook}).  There are also versions of it in the 
computer science literature called the planted multisection model, e.g., McSherry \citeyearpar{McSherry}, and in the community detection literature, e.g., Copic, Jackson and Kirman \citeyearpar{CopicJacksonKirman}.  } 
It consists of a vector $\mathbf{n} = (n_1,\ldots,n_m)$ which captures how many nodes of each type there are (and implicitly, how many types, $m$, there are), and a symmetric $m$-by-$m$ matrix $\mathbf{P}$, whose entries  in $[0,1]$ describe the probabilities of links between various types.  Let $N_k$ be the set of nodes of type $k$, and without loss of generality label nodes so that $\{1, \ldots, n_1\}$ are the nodes of the first type, $\{1+n_1, \ldots, n_1+n_2\}$ are the nodes of the second type, and $N_k=\left\{1+\sum_{i<k} n_i, \ldots, \sum_{i\leq k} n_i\right\}$ are the nodes of the $k$-th type. 
The resulting random network is
captured via its adjacency matrix which is denoted $\mathbf{A}(\mathbf{P},\mathbf{n})$ and is a random variable.
In particular, $\mathbf{A}(\mathbf{P},\mathbf{n})$ is built by letting the entries $A_{ij}$ with $i > j$ be independent Bernoulli random variables with parameter $P_{k \ell}$ if $i \in N_k$ and $j \in N_{\ell}$.   That is, the entry $P_{k \ell}$ captures the probability that an agent of type $k$ links to an agent of type $\ell$. 
%Additionally, we require\footnote{This assumption does not change any of the convergence results substantially, but is clearly the right modeling assumption when a link from $i$ to $j$ means that $i$ has access to the information of $j$.} $A_{ii} = 1$ for all $i$. 
We then fill in the remaining entries of $\mathbf{A}$ by symmetry: $A_{ij} = A_{ji}$. 

Here are some special cases of the model.   

If $P_{k\ell}=p$ for all $k,\ell$, then this is simply an Erd\H{o}s-R\'{e}nyi random network.  

The random network model of Chung and Lu \citeyearpar{chunglu} is the special case where the only heterogeneity in liking is induced by expected degrees. In particular, each type has an expected degree $w_k$, and $P_{k\ell} = w_k w_\ell / W$ where $W=\sum_k n_k w_k$. Thus, the matrix $\mathbb{P}$ is reduced from having $m(m+1)/2$ degrees of freedom to having $m$, namely the expected degrees of the types.\footnote{Of course, if there are as many possible expected degrees as agents, then $\mathbf{P}$ is the same size as the adjacency matrix of the network, which makes the model less tractable than when there are a few types; in empirical settings, usually the number of permitted expected degrees is small compared with the size of the network.}

A spatial model is one where each type of node has a parameter $\theta_k\in \mathbb{R}^m$ for some $m$ that describes it, and 
$P_{k\ell}  = f(\Delta(\theta_k, \theta_\ell))$ where $f$ is a decreasing function and $\Delta$ is Euclidean distance.  This the probability that nodes link to each other is a function of how similar their types are. 

\subsubsection{The Islands Model} \label{sec:islands}

Another special case of the model that we discuss in some of the results below is an {\sl islands model}. There we assume that all $n_i$ are equal, so that islands are equal-sized; we set $P_{kk}= p_s$ for all $k$ and $P_{k\ell}=p_d$ for all $k\neq \ell$.  Thus,
nodes of the same type connect to each other with one probability, and nodes of different types connect to each other with another probability.\footnote{See Currarini, Jackson and Pin \citeyearpar{CurrariniJacksonPin} and Copic, Jackson and Kirman \citeyearpar{CopicJacksonKirman} for illustrations and applications of such a model.}

\medskip

These examples are only a few of the possibilities, and clearly one can consider combinations of these variations, and other considerations such as
special cases where linking probabilities are built on some hierarchy, etc.   

\subsubsection{Remarks on When the Multi-Type Random Networks Model is Useful}

These examples give an idea of how rich the multi-type random networks model is. However, as with any model, it is most pointed in its predictions when we obtain a significant reduction in the dimensionality of the problem.  In particular, for our main results involving representative agents to be most useful, it is helpful for there not to be too many types or, failing that, for the interaction between types to be described by only a few parameters.
Effectively, the results reduce the problem of working with a network of $n$ individuals to a simpler problem of working with $m$ types.  If there are as many types as individuals, 
then clearly that will be unhelpful without additional assumptions; however, a good deal of explanatory power comes out from looking at just a few types, to the extent that a few types capture most of the important variation in the data.  
As we will see from our look at the data, very simple definitions of types have substantial explanatory power.

\subsection{Communication and Learning Processes}

We now describe three different processes communication/learning that we consider.   
As we shall see, there will be some differences in how these are affected by homophily.

\subsubsection{Shortest-Path Communication}

A shortest-path communication process is any process where the time for communication to occur between two nodes (however ``communnication'' is defined) is proportional to the length of the shortest path between the two nodes.\footnote{Standard network definitions, such as shortest path, are omitted.  See Jackson \citeyearpar{JacksonBook} for background definitions.}    
In a connected network, this applies to broadcast processes, where nodes communicate to all neighboring nodes in each period, or to processes where the network is explicitly navigated by a traveler using some sort of addressing system.  This applies to some social and many physical and electronic transmission processes.

\subsubsection{A Repeated Updating Learning Model} 
\label{sec:networkAndUpdating}

The second process that we examine is based on a model first discussed by French \citeyearpar{French} and Harary \citeyearpar{Harary}, and articulated in a more general form by DeGroot \citeyearpar{DeGroot}.  

Given a network $\mathbf{A}$, let $\mathbf{T}(\mathbf{A})$ be defined by $T_{ij}(\mathbf{A}) =  A_{ij}/d_{i}$. 
Beginning with some initial belief vector $\mathbf{b}(0)\in [0,1]^n$, let $$\mathbf{b}(t) = \mathbf{T}(\mathbf{A}) \mathbf{b}(t-1)$$
 for all $t \geq 1$. That is, agents form today's beliefs by taking the average of neighbors' beliefs yesterday, where an agent can be his own neighbor. 
It is immediate that then  
$$\mathbf{b}(t) = \mathbf{T}(\mathbf{A})^t \mathbf{b}(0).$$

If the initial beliefs $\mathbf{b}(0)$ are independent and identically distributed draws from normal distributions around a common mean then the linear updating rule at $t=1$ corresponds to Bayesian updating with certain priors about signal precisions as discussed by DeMarzo, Vayanos, and Zwiebel \citeyearpar{PersuasionBias}.  The behavioral aspect of the model concerns times after the first round of updating.  Here, it is no longer Bayesian to update using a weighted-average rule, but due to the overwhelming complexity of the Bayesian calculation, we assume agents continue using the simple averaging rule in later periods, too.  More discussion of this assumption can be found in DeMarzo, Vayanos, and Zwiebel \citeyearpar{PersuasionBias}.

Beyond the interpretation of updating signals and ``learning'', the linear updating model can also be interpreted as a model that captures behaviors where agents adjust their behaviors to match the average of their neighbors' choices.  In particular, it can be interpreted as myopic best-response updating in a game. Suppose that agents have to choose each period a variable $b_i$, which captures their behaviors, for instance which dialect of a language they speak, and the dialects correspond to points in $[0,1]$. The cost to $i$ of communicating with $j$ is $(b_i-b_j)^2$.  If each agent communicates with his neighbors according to $\mathbf{A}$, then the best response mapping is given by the linear updating rule.

If $\mathbf{T}(\mathbf{A})$ is not connected, then it suffices to consider the asymptotic behavior of each connected component to understand the full dynamics of the process.\footnote{If the communication network is directed then convergence requires some aperiodicity in the cycles of the network and works with a different segmentation into components, but still holds quite generally,
as discussed in Golub and Jackson \citeyearpar{GolubJackson2007}.} 
Thus, we assume from now on that $\mathbf{T}(\mathbf{A})$ is connected, and define $\mathbf{T}(\mathbf{A})^\infty = \lim_{t \to \infty} \mathbf{T}(\mathbf{A})^t$.  

\begin{lemma}
\label{lemma-limit}
If $\mathbf{A}$ is connected, then $\mathbf{T}(\mathbf{A})^t$ converges to a limit $\mathbf{T}(\mathbf{A})^\infty$ such that $(\mathbf{T}(\mathbf{A})^\infty)_{ij}= \frac{d_i(\mathbf{A})}{D(\mathbf{A})}$.
\end{lemma}

Lemma \ref{lemma-limit}
follows from standard results on Markov chains (e.g., see Golub and Jackson \citeyearpar{GolubJackson2007} and Jackson \citeyearpar{JacksonBook} for details and background)
and implies that for any given initial vector of beliefs $\mathbf{b}(0)$, 
the limiting belief 
$$\lim_t \mathbf{b}(t) = \mathbf{T}^\infty \mathbf{b}(0)=(b,b,\ldots,b) {\rm \ \ \  where \ \ \ } b =\sum_i \frac{b_i(0) d_i(\mathbf{A})}{D(\mathbf{A})}.$$  
Thus, the relative influence that an agent has over the final beliefs is his or her relative degree.

\subsubsection{Random Walks}
\label{sec:randomwalks}

The third process that we study is a random walk on a network. This is a process where a particle starts at some node and hops to any of its neighbors with equal probability at each step. One example to think of is that of a college student who is viewing Facebook profiles; at each step, she clicks on a random friend of the person whose profile she is currently viewing. Another example is the Google model of a surfer who randomly clicks on links as he navigates the World Wide Web.\footnote{Of course, we might think of him being biased toward following certain out-links from a given web page; this could be modeled by using a nonuniform random walk; i.e., one in which all nonzero entries in a given row of $\mathbf{T}$ are not the same. We suspect that our conclusions would simply be modified by weighting factors, but the symmetry of the simpler case is handy in our proofs.} 
Here, the particle starts at some location and transitions from node $i$ to node $j$ with probability $T_{ij}$.  The question is how long it takes to reach the steady state
distribution on location.   While this is a fairly specific process, and may not capture as many applications as the previous two processes, it has figured prominently in the literature Markov processes and random graphs and
so it is a very useful benchmark.  

Just as in the case of the linear updating learning model, Lemma \ref{lemma-limit} implies that if $\mathbf{A}$ is connected, then ${\mathbf{T}^t}$ converges
and then the limit distribution of the random walk is to be at node $i$ with probability  $\frac{d_i(\mathbf{A})}{D(\mathbf{A})}$ regardless of the starting position of the random walk.
So, the limiting distribution of the time that the walk spends at a given node is proportional to its degree.

\subsection{Consensus Time and Mixing Time}

We now present ways of measuring the speed at which the above-defined processes operate on a given network.   The different processes suggest different measures.   

First, shortest-path based processes have an obvious measure of speed, which is simply the average shortest path length in the network, or if one is worried about the longest time it could take to pass from some node to some other node, then the diameter of the network.  These are standard notions, so there is no need to develop any special measure for such processes.
  
The other two processes require measures of timing/distance that are more tailored to them.  We now discuss each in turn.

\subsubsection{Distances between Vectors}

There are two distance measures that we focus on in measuring convergence.
   
The first is a standard weighted squared deviation distance.  Given two vectors of beliefs $v$ and $u$ let
$$  \Vert \mathbf{v} - \mathbf{u} \Vert ^2_{\mathbf{w}} = \sum_i w_i (v_i - u_i)^2. $$

In applying this, we will be interested in the differences
between beliefs at time $t$ and their limit:
$$ \Vert \mathbf{T} (\mathbf{A})^t\mathbf{b}- \mathbf{T} (\mathbf{A})^\infty\mathbf{b} \Vert ^2_{\mathbf{w}}.$$
It will be useful to use weights, ${\mathbf{w}}$, that are the influences of the agents $\mathbf{s}(\mathbf{A})$,
where
$$\mathbf{s}(\mathbf{A})=\left( \frac{d_1(\mathbf{A})}{D(\mathbf{A})}, \ldots, \frac{d_n(\mathbf{A})}{D(\mathbf{A})} \right).$$
The distance
$$ \Vert \mathbf{T}(\mathbf{A})^t\mathbf{b}- \mathbf{T}(\mathbf{A})^\infty\mathbf{b} \Vert ^2_{\mathbf{s}(\mathbf{A})}$$ 
examines the squared difference between agents' current beliefs and their limit beliefs.  The distance is weighted by the agents' degrees which gives more weight to relatively more influential 
agents.  
This quantity has a fairly simple interpretation.  Consider the following experiment: agents start with beliefs $\mathbf{b}$; at time $t$, an agent is sampled uniformly at random. We imagine that he asks a random neighbor for his opinion: i.e. one of his neighbors is sampled uniformly at random; we  record the square of this neighbor's deviation from consensus beliefs. The expectation of this variable under this experiment is the distance defined above. In other words, sampling each agent in proportion to his degree captures the deviation from consensus of an opinion sent at time $t$ across a randomly chosen link in the network.

The following lemma shows an obvious relationship between a straight averaging of
the squared deviations in beliefs and this weighted averaging.

\begin{lemma} \label{lem:relatingDeltaAndD} Let $\mathbf{e}=(1,\ldots, 1)$.  Then
\begin{align*} \frac{\bar{d}(\mathbf{A})}{d_{\max}(\mathbf{A})}  \Vert \mathbf{T}(\mathbf{A})^t\mathbf{b}- \mathbf{T}(\mathbf{A})^\infty\mathbf{b} \Vert ^2_{s(\mathbf{A})} &\leq 
 \Vert \mathbf{T}(\mathbf{A})^t\mathbf{b}- \mathbf{T}(\mathbf{A})^\infty\mathbf{b} \Vert ^2_{\mathbf{e}/n}   \\ &\leq \frac{\bar{d}(\mathbf{A})}{d_{\min}(\mathbf{A})} \Vert 
\mathbf{T}(\mathbf{A})^t\mathbf{b}- \mathbf{T}(\mathbf{A})^\infty\mathbf{b} \Vert ^2_{s(\mathbf{A})}. \end{align*}
\end{lemma}
Thus, for graphs where the highest- and lowest-degree agents have degrees not too different from the average, as in some of the networks we will be concerned with, whether or not we weight mean square deviation from consensus by degree will not make a large difference.  Even in networks where there are large deviations in degree, if the number of deviant nodes is bounded, then a direct variation of Lemma \ref{lem:relatingDeltaAndD} implies that the two notions are still close. Given that converting between these two notions only requires computing some bounds which will usually be good but which will depend on the application, we work with the degree-weighted version of the deviation measure as it has nicer mathematical properties and more intuitive
relationships to the network structure.

\bigskip

The other distance measure that we will work with is the total variation metric, where the
distance between a vector $\mathbf{v}$ and another vector $\mathbf{u}$ is
$$  \Vert \mathbf{v} - \mathbf{u} \Vert ^{TV} = \frac{1}{2}\sum_i |v_i - u_i|. $$

We will be applying this in cases where $\mathbf{v}$ and $\mathbf{u}$ are probability measures, so that $\mathbf{v} \geq \mathbf{0}$, $\mathbf{u} \geq \mathbf{0}$, and $\sum_i v_i=\sum_i u_i=1$. Then it is straightforward to see that 
$$  \Vert \mathbf{v} - \mathbf{u} \Vert ^{TV} = \frac{1}{2}\sum_i |v_i - u_i|= \max_{C\subset N} \left|\sum_{i\in C} v_i -\sum_{i\in C} u_i \right|, $$
so the total variation metric keeps track of the maximal difference in the probability that two measures assign to some set.

\subsubsection{Consensus Time and the Linear Updating Model}

A central question in the linear updating model is the rate at which beliefs of the society converge to their consensus limit.
We define the consensus time as follows.

\begin{definition} 
The \emph{consensus time} to $\varepsilon > 0$ of the network $\mathbf{A}$ is
$$ \CT(\varepsilon;\mathbf{A}) = \sup_{\mathbf{b} \in [0,1]^n} \min \{t :  \Vert \mathbf{T}(\mathbf{A})^t\mathbf{b}- \mathbf{T}(\mathbf{A})^\infty\mathbf{b} \Vert ^2_{\mathbf{s}(\mathbf{A})}< \varepsilon \}.$$ 
\end{definition}

The need to consider different potential starting belief vectors $\mathbf{b}$ is clear, as if one starts with $b_i(0)=b_j(0)$ for all $i$ and $j$ then consensus is reached instantly.  Thus, the ``worst
case'' $\mathbf{b}$ will generally have beliefs that differ across types and is useful as a benchmark measure of how homophily matters; taking the supremum in this way is standard in defining convergence times (e.g., see 
Montenegro and Tetali \citeyearpar{montenegrotetali}).
%\footnote{Numerical experiments available on request show that as along as we start with sufficient heterogeneity 
%in beliefs, the rate of convergence will be the same for a wide variety of starting beliefs.} 

Since $\mathbf{T}$ is a contraction under the distance measure  (a standard fact about reversible Markov chains), once the mean-square deviation is below $\varepsilon$, it can never go above it again. Thus, the definition is equivalent to letting $\CT(\varepsilon;\mathbf{A})$ be the earliest time such that deviation from consensus is small forever after.

\subsubsection{Mixing Time and Random Walks}

There are many definitions measuring the distance of a random process to its limit that
come from the 
literature on Markov chains (e.g., see Montenegro and Tetali \citeyearpar{montenegrotetali}), and the following definition is among the most common.  Let $\mathbf{e}_i$ be the unit vector with an entry of $1$ in the $i$-th entry and 0's elsewhere.  

\begin{definition} 
The \emph{mixing time} to $\varepsilon > 0$ of a network $\mathbf{A}$ is  
$$ \MT(\varepsilon;\mathbf{A}) = \sup_{i} \min \{t :  \Vert \mathbf{e}_i\mathbf{T}(\mathbf{A})^t- \mathbf{e}_i\mathbf{T}(\mathbf{A})^\infty \Vert ^{TV} < \varepsilon \}.$$ 
\end{definition}

Mixing time keeps track of how different the probability across states is after $t$ periods compared to the limiting distribution.  Taking the supremum over different starting states is equivalent to considering all possible starting distributions.

\subsubsection{The Relation of Consensus and Mixing Time}

The consensus and mixing times have a close relationship intuitively, as both depend on how quickly $\mathbf{T}^t$ approaches its limit.  
The main difference is that consensus time works with mean-squared deviations while mixing time works with the sum of absolute values of the deviations.  In terms of the mathematics, the difference between them is simply the difference
between the $\ell^2$ and $\ell^1$ norms as well as a difference in the normalizing constant. 

Just as an illustration, consider the distance between $(1,0,0,\ldots)$ and $(1/n,1/n,1/n,\ldots)$. 
If we think of these as probability measures, then they are quite different as one is a Dirac measure and the other is a uniform distribution.   In contrast, if these are behaviors or beliefs, then only one agent in the society is deviating substantially from the limiting behavior or beliefs.  Thus depending on the application, one might or might not want to consider these to be close or far apart.   Under the $\ell^2$ norm used in calculating consensus time,
these are close to each other, while under the $\ell^1$ norm used in calculating mixing time, they are quite far apart.

The mixing time approach is most natural and standard in the setting of Markov chains, where distributions are important, and conensus time is a natural measure in the setting of linear updating models. We present results on both, and shall see that despite the differences, they will behave similarly in our setting under the right normalizations.  As we shall see also in the empirical section (see Figure \ref{fig:ctversusmt}), consensus time and mixing time will essentially coincide in the high school data.

\subsubsection{Asymptotics}

In some cases, we consider what happens as $n$ grows.   This is natural as there are many properties of random graphs that can be deduced to hold
almost surely for ``large'' networks, but that are hard to express in any meaningful way for a small random graph where any possible configuration has a nontrivial probability of arising.  
In such cases, then there is some question as to the appropriate choice of $\varepsilon$.   

Given the steady state distribution ${\mathbf{s(A)}}=\left(\frac{d_1(\mathbf{A})}{D(\mathbf{A})}, \ldots, \frac{d_n(\mathbf{A})}{D(\mathbf{A})}\right)$,
if the $d_i(\mathbf{A})$'s are growing at roughly the same speed (a condition in some of the results below), then the entries of $\mathbf{s(\mathbf{A})}$ will be of order $1/n$. As such, a natural benchmark is to examine $\CT(\gamma/n^2; \mathbf{A})$  (given the squaring in the norm) and $\MT(\gamma/n; \mathbf{A})$ for some fixed $\gamma>0$.

\section{The Speed of Communication}

We now present our main conceptual conclusions about the speed of communication and learning for the various processes and discuss the contrasts across 
different sorts of communication.   We then come back to the main technical contributions in the next section, which seem to be of some independent interest.

\subsection{Shortest-Path Communication}

Consider the multi-type random network $(\mathbf{n},\mathbf{P})$ with an associated 
number of nodes $n$ and let $d_k(\mathbf{n},\mathbf{P})= \sum_{k'} P_{kk'} n_{k'}$ indicate the expected degree of a node in group $k$,  
$\widetilde{d}(\mathbf{n},\mathbf{P})=\sum_k (d_k(\mathbf{n},\mathbf{P}))^2 n_k/D(\mathbf{n},\mathbf{P})$ be the second order average degree in society,\footnote{Note that if 
the average degree $d_k(\mathbf{n},\mathbf{P})$ is the same across groups, then this is just the average degree.} $$D(\mathbf{n},\mathbf{P}) = \sum_k d_k(\mathbf{n},\mathbf{P}) n_k$$ 
be the total expected degree, and $p(\mathbf{n},\mathbf{P})=D(\mathbf{n},\mathbf{P})/n$ be the average probability of a link.
Suppose that 
\begin{itemize}
\item[(i)] there exists $M<\infty$ such that $ \max_{k,k'} d_k(\mathbf{n},\mathbf{P})/d_{k'}(\mathbf{n},\mathbf{P})<M$,
\item[(ii)] $\widetilde{d}(\mathbf{n},\mathbf{P})\geq (1+\varepsilon )\log(n)$ for some $\varepsilon>0$, 
\item[(iii)]  $ \log( \widetilde{d}(\mathbf{n},\mathbf{P})) / \log(n)\rightarrow 0$, and
\item[(iv)] there exists $\varepsilon >0$ such that $\min_{kk'} P_{kk'} / p(\mathbf{n},\mathbf{P})>\varepsilon$.
\end{itemize}

These conditions admit many cases of interest and can be understood as follows: (i) implies that there is not a divergence in the expected degree across groups; (ii) ensures that the average degree grows with $n$ fast enough so that the network becomes connected with a probability going to 1, so that the network will not have isolated components across which communication is impossible;  (iii) implies that average degree grows more slowly than $n$, as otherwise the shortest path degenerates to being of length 1 or 2 and does not match most empirical applications; and (iv) that there is some lower bound on the probability of a link between groups relative to the overall probability of links in the network.  This last condition ensures that groups do not become so homophilous that the network becomes disconnected.  These conditions still allow for substantial homophily.  For instance, if $n_k$ is on the order of $d(\mathbf{n},\mathbf{P})$, then this still allows the probability of a link within same group to even become infinite relative to the probability across links, so that arbitrarily high levels of homophily are permitted.  
 
\begin{theorem}[Jackson \citeyearpar{JacksonDiameter}]\footnote{This result holds for a more general random network model, and is specialized to the multi-type random network model considered here for this statement.}\label{thm:avgdist}
If the random network process $(\mathbf{n},\mathbf{P})$ satisfies (i)-(iv) then, asymptotically almost surely in $n$, $\mathbf{A}(\mathbf{n},\mathbf{P})$ is connected; the average distance between nodes is $(1+o(1))\log(n)/\log(\widetilde{d}(\mathbf{n},\mathbf{P}))$; and the 
diameter of the largest component is $\Theta (\log(n)/\log(\widetilde{d}(\mathbf{n},\mathbf{P})))$. \label{thm:shortestpaths}
\end{theorem}

Theorem \ref{thm:avgdist} tells us that although homophily can change the basic structure of a network, it does not affect the average shortest-path distance between nodes in the network.   Moreover, we have a precise expression for that average distance which is the same as it is in an Erd\"os-R\'enyi random network with the same average degree.  
Effectively, as we increase the homophily, we increase the density of links within a group but decrease the number of links between groups.  The result is perhaps somewhat surprising in showing that these two effects perfectly balance each other to keep average path length unchanged.  
The intuition behind the theorem can be understood in the following manner.  Suppose that every node had a degree of $d$ and that the network was a tree.  Then the $k$-step neighborhood of a node would capture roughly $d^k$ nodes.  Setting
this equal to $n$ leads to a distance of $k=\log(n)/\log(d)$ to reach all nodes, and given the exponential expansion, this would also be the average distance.   The theorem shows that this is exactly how the average distance behaves even when the network is not a tree, even when we noise up the network so that nodes do not all have the same degree, and even when we add substantial homophily to the network.  
In proving this, there are two critical parts:  first, the randomness of the nodes' degrees does not substantially alter the calculation (even if a power law distribution is admitted in 
expected degrees); and second, even though homophily may alter the structure of the network, the shortest paths branching out from a given node are not much altered by homophily.  The homophily affects which nodes are likely to be closer or further, but not the average distance.  

\begin{corollary}
Consider a process that has an expected communication time equal to the average distance
between nodes.  If it is run on two different random network formation sequences
satisfying (i) to (iv) that have the same second order average degree as a function of $n$,
then the ratio of the expected communication times on the two different random network
sequences goes to 1, asymptotically almost surely.
\end{corollary}

The above results tell us that average distance is not affected by homophily, and
diameter is affected only up to a fixed finite factor,  provided there is some minimal 
level of inter-group connectivity.   Thus average-path based communication processes are not affected by homophily but are affected by the link density in a society.

\subsection{Markovian Processes: Linear Updating and the Random Walk}

When we turn to the other forms of communication, there are substantial effects of homophily.  It is not simply 
average distance that matters, but the \emph{relative numbers} of paths between different nodes that matters.

To state the conceptual results most cleanly, we specialize to the islands model.  The results extend to the more general multi-type random networks; those extensions require a somewhat longer exposition and so are discussed in the next section.

In the context of the islands model, let us define two measures of homophily. The (unnormalized) homophily is defined as $$H = \frac{p_s}{p}$$
and captures how much more probable a link to a node of one's own type is compared to other types.   This varies between $0$ and $m$, the number of islands.
If a node only links to same-type nodes, then the average linking probability $p$ becomes $p_s/m$ and so $H=m$, while if a node only links to nodes of other types, then
$p_s=0$ and so $H=0$.  
We can also normalize the measure by dividing by the number of islands $m$; the normalized homophily is thus defined as
$$h = \frac{p_s}{mp}.$$
Thus, $h$ is the fraction of a node's links that are expected to be to agents of the same type.

If we index a sequence of societies by
their cardinalities $n$,
then the following theorem summarizes the main conclusions of how homophily affects consensus and mixing times. The details and proof in a more general setting appear in the next section.

\begin{theorem}
\label{thm:islandslimits}
In the equal-sized islands model, if $p(n)n/\log^2 (n)\rightarrow\infty$ and $h(n)$ is bounded away from 1,
then for any $\delta<1$, high enough $n$ ensures that the following is true with arbitrarily high probability:
$$
\frac{(1-\delta)\log(n)}{2\log(\frac{m-1}{H-1})}
\leq  \CT(\gamma/n^2; \mathbf{A}(\mathbf{P}, \mathbf{n})) \leq
\frac{(1+\delta)\log(n)}{\log(\frac{m-1}{H-1})}
$$
and 
$$
\frac{(1-\delta)\log(n)}{\log(\frac{m-1}{H-1})}
\leq  \MT(\gamma/n; \mathbf{A}(\mathbf{P}, \mathbf{n})) \leq
\frac{\frac{3}{2}(1+\delta)\log(n)}{\log(\frac{m-1}{H-1})}
.$$
\end{theorem}

Theorem \ref{thm:islandslimits} provides us with a precise relationship between homophily and 
consensus and mixing times.  
As homophily increases, both consensus time and mixing time increase.  
In particular, both the
consensus time $ \CT(\gamma/n^2;  \mathbf{A})$ and the mixing time
$ \MT(\gamma/n; \mathbf{A})$
are proportional (up to a fixed factor)
to 
$$
\frac{\log(n)}{\log(\frac{m-1}{H-1})}
.$$
This is true independently of link density and does not impose any specific requirements about
how many types (islands) there are.
If $m$ becomes large, then this further simplifies and both the
consensus time $ \CT(\gamma/n^2;  \mathbf{A})$ and the mixing time
$ \MT(\gamma/n; \mathbf{A})$
are proportional (up to a fixed factor)
to 
$$
\frac{\log(n)}{\log(1/h)}
.$$
When homophily is low, then marginal increases in homophily has only a small effect. But as homophily grows large ($H/m$ or $h$ closer to $1$), the magnitude of the marginal effect of increased homophily becomes very large.

\subsection{Discussion}

The interesting and intuitive contrast is the comparison between what matters for shortest-path and Markovian communication processes. Comparing Theorems \ref{thm:avgdist} and \ref{thm:islandslimits}, we see that, in the context of communication based on shortest paths, homophily has no effect while average link density is critical. In contrast, when considering Markovian processes, we see that homophily is critical while average link density is irrelevant.  

This is an intuitive difference. With shortest-path communication, by definition, only the average distance between pairs of agents matters; it does not matter who is close and who is far from a given agent. Adding homophily changes network structure on the second dimension -- introducing a certain kind of clustering (see Jackson \citeyearpar{JacksonDiameter} for more on clustering) -- but it does not change the average lengths of the shortest paths branching out from each node. Agents just end up very close to other agents of their own types, who then provide connections to other types. While these last statements are not obvious, Theorem \ref{thm:avgdist} asserts that the various effects interact in just the way needed to keep average distances the same.

With Markovian communication, on the other hand, the time to convergence is determined by how  homophilous the network is, not by the density of its connections, provided that a certain (low) density threshold is met. A network in which the underlying links are formed without discrimination will not have any way to support long-term heterogeneity away from the steady state, even if the network is fairly sparse. On the other hand, a network with clearly defined islands of clustered agents will be able to maintain long-term differences in beliefs or behaviors. Each island will converge to its own metastable state and stay there for a long time, disagreeing with the other islands. Increasing the overall number of links while maintaining the homophily will not speed up convergence. The ratio of same-type connections to different-type connections for a given node will remain essentially the same, and so the network will be able to support the same disagreement based on the self-reinforcing effect within given types.

\section{Relating Communication Times, Second Eigenvalues, and Homophily}

Theorem \ref{thm:islandslimits} is proven via a series of mathematical results which are of some interest in their own right. Again, as outlined in the introduction, the relation of speed of learning to homophily is established by breaking things into two parts:   how speed relates to second eigenvalues, and how 
second eigenvalues relate to homophily.  The key result that unlocks the second part of this puzzle is a representative agent theorem that shows that the second eigenvalue of a 
random network is, asymptotically, only dependent on the underlying probabilities of linking between different types of agents.  The extra noise of which specific agents are linked to which others is essentially irrelevant in a large network. Only the broad patterns of linking across different types are important.  Once we have proved this, we can relatively easily deduce results relating second eigenvalues to homophily, and then complete the picture relating speed to homophily.

\begin{figure}[tp]
\begin{center}
     \includegraphics[height=7in]{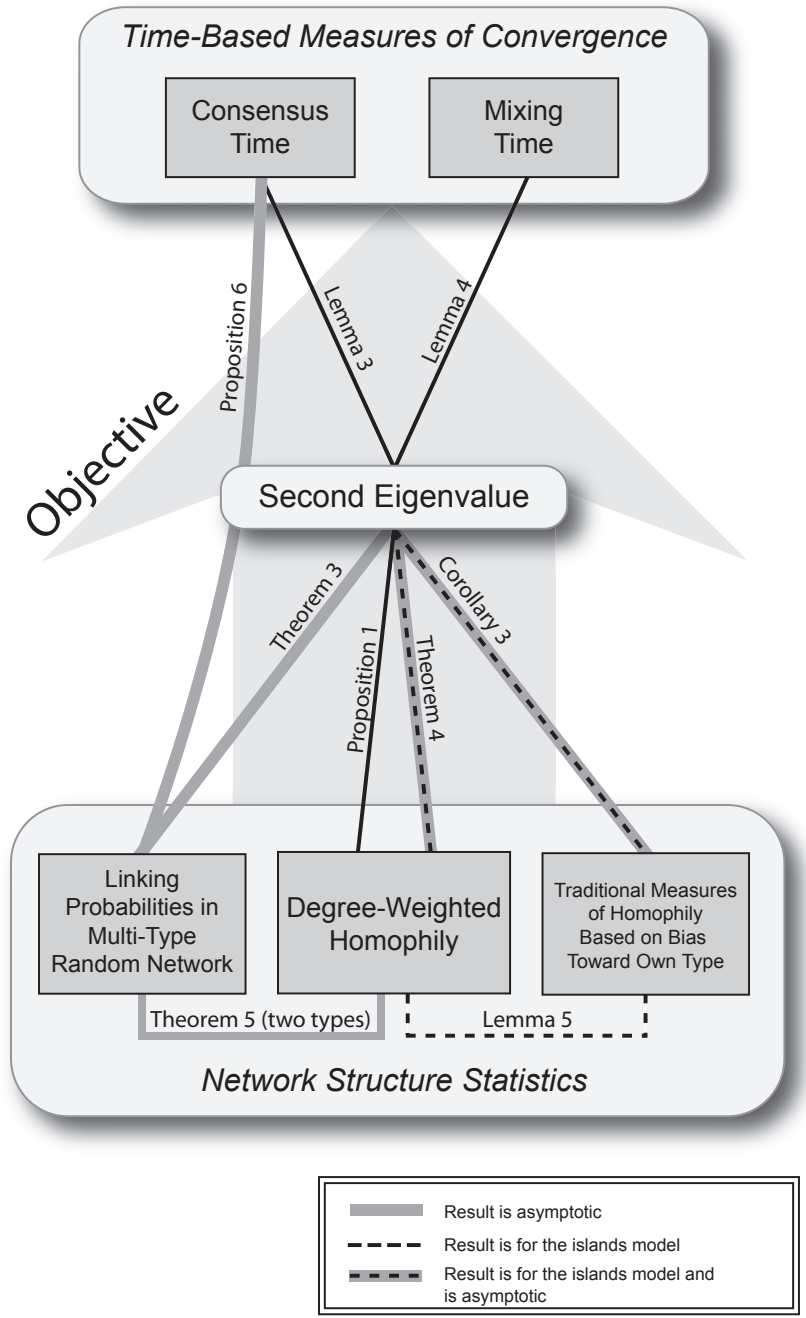}
\end{center}
\caption{The paper's conceptual structure. A line between two quantities indicates that one is used to bound or characterize the other; quantities located lower in the diagram are used to characterize the ones that are higher.}
\label{fig:diagram}
\end{figure}

The outline is summarized in Figure \ref{fig:diagram}. A specific roadmap of the technical results is as follows. 

Measures of convergence speed form the top layer of Figure \ref{fig:diagram} and summary statistics related to large-scale network structure form the bottom layer. The middle layer is the spectral intermediary that allows us to tie everything together. It is well-known that the second eigenvalue of a network's Markov matrix is a good proxy for the convergence speed of linear updating processes. Thus, in relating the top and middle layers of Figure \ref{fig:diagram}, we use standard spectral results from the Markov chain literature to provide upper and lower bounds on mixing and consensus times. The results on mixing time, which is a central concept in Markov chain theory that captures how long it takes a process on a network to become random, are completely standard. We also define a notion of consensus time, which is essentially the time required for the mean squared deviation from consensus beliefs to get small. Bounding this quantity requires adapting standard results in a straightforward way (Lemma \ref{lem:ConsensusAndEigenvalue}) but it turns out that in the multi-type random graph setting, tighter bounds than usual can be obtained for it (Proposition \ref{prop:homophilousConvergenceBounds}).

The main novel technical work of the paper concerns the relationship between the middle and bottom layers of the figure. First, we prove a general result, Theorem \ref{thm:representativeAgent}, which shows that in large multi-type random network, the study of the second eigenvalue of the entire network can be reduced to a computation based on a representative agent matrix which contains only one agent for each type. Building on this, we  relate the second eigenvalue to more concrete measures of homophily. One that turns out to be particularly well suited to the study of eigenvalues, and hence of convergence, is a new quantity called degree-weighted homophily (DWH). This quantity measures the relative advantage of same-type links over different-type links, but does so in a way that takes into account different degrees and group sizes. Proposition \ref{prop:eigenvalueBound} shows that, in arbitrary networks, this quantity always provides a lower bound on second eigenvalue, and hence consensus time. These results already entail that homophily slows learning and provide general tools for studying the relationship in arbitrary multi-type networks. For more concrete characterizations in an important special case, we turn to networks in which agents split into equally sized ``islands'', where each island is a different type. Agents only discriminate based on whether someone else is inside or outside of their own islands. In this case, for large networks, we can exactly characterize second eigenvalues for large networks. These results can be stated in terms of a close relative of DWH (Theorem \ref{thm:islandslimits}) and in terms of more traditional unweighted measures of homophily (Corollary \ref{cor:traditionalhomophilies}). 

Putting the transitions between the layers together, we end up with a tight relationship for large island networks between homophily and the speed of learning, which then leads to the conclusions described above. These are summarized in Theorem \ref{thm:islandslimits}.

\subsection{Relating Consensus and Mixing Times to Second Eigenvalues}

We first state results that give fairly precise bounds on how
the consensus time and mixing time of a matrix depends on its second eigenvalue. For any stochastic matrix $\mathbf{T}$, let $1=\lambda_1(\mathbf{T}),\ldots,\lambda_n(\mathbf{T})$ be its eigenvalues sorted by magnitude in decreasing order.

\subsubsection{Consensus Time}

\begin{lemma} 
\label{lem:ConsensusAndEigenvalue}
Assume $\mathbf{A}$ is connected, let $\lambda_2(\mathbf{T}(\mathbf{A}))$ be the second largest eigenvalue
in magnitude of $\mathbf{T}(\mathbf{A})$, and let $\mathbf{s}$ be the (unique) steady-state distribution, with $\min_i s_i=\underline{s}$. If $\lambda_2(\mathbf{T}) \neq 0$, then for any $0 < \varepsilon \leq 1$:
$$\left \lfloor \frac{\log (1/4\varepsilon) - \log(1/\underline{s})}{2\log (1/| \lambda_2(\mathbf{T})|)} \right \rfloor \leq \CT(\varepsilon; \mathbf{A}) \leq \left \lceil \frac{\log (1/\varepsilon)}{2\log (1/| \lambda_2(\mathbf{T})|)} \right \rceil.$$ If $\lambda_2(\mathbf{T}) = 0$, then for every $0 < \varepsilon < 1$ we have $\CT(\varepsilon; \mathbf{A}) = 1$.
\end{lemma}

If $\varepsilon$ is fairly small, then the bounds in the lemma are close to each other and so we have 
a quite precise characterization in terms of the spectrum of the underlying social network. 

The proof of this result follows fairly standard techniques from the spectral literature and the proof appears in the appendix.

The lower bound in Lemma \ref{lem:ConsensusAndEigenvalue} includes a term $\log(1/\underline{s})$ which can grow as $n$ grows.  We improve on the lower bound in 
Proposition \ref{prop:homophilousConvergenceBounds} in the appendix, where we take advantage of the random graph structure to obtain a lower bound that does not depend on $\underline{s}$ or $n$.

\subsubsection{Mixing Time}

Next, let us consider mixing time.  Again, we can derive bounds based on the second eigenvalue.  
In this case, there is a difference that reflects the difference in the norms associated with these measures. 

The following lemma is adapted from Montenegro and Tetali \citeyearpar{montenegrotetali} (see their Section 2.4: ``Does Reversibility Matter'').  

\begin{lemma}
\label{mixing} [Montenegro and Tetali \citeyearpar{montenegrotetali}]
Assume $\mathbf{A}$ is connected. Let $\lambda_2(\mathbf{T})$ be the second largest eigenvalue
in magnitude of $\mathbf{T}$, and $\mathbf{s}$ be the (unique) steady-state distribution, with $\min_i s_i=\underline{s}$. If $\lambda_2(\mathbf{T}) \neq 0$, then for any
$0 < \varepsilon\leq 1$:
$$
\frac{\log(\frac{1}{2\varepsilon})}{\log(\frac{1}{|\lambda_2(\mathbf{T})|})}
\leq  \MT(\varepsilon; \mathbf{A}) \leq
\frac{\log(\frac{1}{2\varepsilon})+ \log(1/\underline{s})/2}{\log(\frac{1}{|\lambda_2(\mathbf{T})|})}
.$$ If $\lambda_2(\mathbf{T}) = 0$, then for every $0 < \varepsilon < 1$ we have $\CT(\varepsilon; \mathbf{A}) = 1$.
\label{lem:mixingBounds}
\end{lemma}

\subsection{Relating Second Eigenvalues to Network Structure}

\subsubsection{A Representative Agent Theorem}
 
We now present our main technical result, a ``representative-agent'' theorem that allows us to analyze the convergence of a multi-type random graph by studying a much smaller graph in which there is only one node for each \emph{type} of agent. 
We show that under some conditions on the minimum expected degree, the second eigenvalue of most any realized multi-type random graph converges in probability to the second eigenvalue of this representative-agent matrix.  This result is useful for dramatically simplifying computations of approximate consensus times and mixing times, both in theoretical results and in empirical settings, as now the random second eigenvalue can be accurately predicted knowing only the relative probabilities of connections across different types, as opposed to anything about the precise realization of the random network.

Recalling the notation from Section \ref{sec:multitype}, 
let $d_{k\ell}(\mathbf{P},\mathbf{n}) = n_\ell P_{k\ell}$ the expected number of links that a node of type $k$ will 
have with nodes of type $\ell$ and let
$d_k(\mathbf{P},\mathbf{n}) = \sum_\ell d_{k\ell}(\mathbf{P},\mathbf{n})$ be the expected degree of a node of 
type $k$.  
Let $\mathbf{Q}(\mathbf{P}, \mathbf{n})$ be a matrix of the same dimensions as $\mathbf{P}$ with entries 
$$Q_{k\ell} = \frac{d_{k\ell}(\mathbf{P},\mathbf{n})}{d_{k}(\mathbf{P},\mathbf{n})}.$$
So, $Q_{k\ell}$ is the expected relative fraction of links that a node of type $k$ will have
with nodes of type $\ell$. 
This simplifies things in two respects relative to the realized random network.  First, it works with groups rather than individual nodes, and second, it works with expected fractions rather than realized values.\footnote{There is one technical issue which does not substantially affect any results but which deserves some comment. If one is thinking of a multi-type random network as describing the relationships relevant for a boundedly rational process of belief updating, then the self-weights assumed in our formulation of the model are unnatural. In particular, if we think of a link from $i$ to $j$ as capturing the fact that $i$ has access to the belief of $j$, then all nodes should have self-links; in the model as we have described it, a node of type $k$ has a self link with probability only $P_{kk}$. It turns out that formulating the model so that all nodes always have self-links does not change any of the results presented in this section, except for a small change in asymptotic rates of convergence in Theorem \ref{thm:representativeAgent} and its consequences. This is because the spectral norm of the difference between the matrix we work with in the proof and the matrix with self-links decays to $0$, assuming that minimum degree is unbounded. The same is true if we forbid self-links, which is natural in the dialect-choice game of Section \ref{sec:networkAndUpdating} or the Facebook random walk of Section \ref{sec:randomwalks}. In short, the results are not sensitive to how we model self links, provided that everyone treats all neighbors, including oneself, equally.}

\begin{theorem} Consider a multi-type random network described by $(\mathbf{P},\mathbf{n})$. For any $\delta > 0$ there exists $K$ such that if  $\min_k d_k(\mathbf{P},\mathbf{n}) > K \log^2 n$ 
then 
\begin{equation} \label{eqn:representativeAgentBound} \left| \lambda_2(\mathbf{T}(\mathbf{A}(\mathbf{P}, \mathbf{n}))) -  \lambda_2(\mathbf{Q}(\mathbf{P}, \mathbf{n})) \right| \leq  \delta, \end{equation} 
with probability at least $1-\delta$.
\label{thm:representativeAgent} \end{theorem}

Theorem \ref{thm:representativeAgent} is a law of large numbers for spectra of multi-type random graphs. Such techniques are a central tool in the random graphs literature; they show that various important properties of random graphs converge to their expectations, which shows that these locally haphazard objects have very precise global structure.   The closest antecedent to this particular theorem is by Chung, Lu and Vu \citeyearpar{ChungLuVu} for Markov or Laplacian matrices of networks without homophily. This theorem is the first of its kind to apply to these matrices in a model that allows homophily and the associated heterogeneities in linking probabilities.\footnote{The result has a superficial similarity to Theorem 10 in McSherry \citeyearpar{McSherry}, which is based on the work of F\"{u}redi and Koml\'{o}s \citeyearpar{FurediKomlos} and Alon, Krivelevich, and Vu \citeyearpar{AlonKrivelevichVu}. These results show that the spectra of certain random matrices are close to the spectra of their expectations. However, in those papers, the matrices of interest are adjacency matrices, whose entries are independent, and this independence is crucial to the arguments. Since the matrix of interest here is the updating matrix $\mathbf{T}(\mathbf{A})$, whose entries are not independent (because each row is normalized), the same techniques do not go through. Instead, we build on work of Chung, Lu, and Vu \citeyearpar{ChungLuVu} and on the other papers just cited to prove a theorem that has a similar flavor but in this different and more complex setting.}  We employ a similar strategy of proof, which relies on decomposing the random matrix representing our graph into two pieces: an ``orderly'' piece whose entries are given by linking probabilities between nodes of various types, and a noisy piece due to the randomness of the actual links. By bounding the spectral norm of the noise, we show that, asymptotically, the second eigenvalue of the orderly part is, with high probability, very close to the second eigenvalue of the random matrix of interest.  Then we note that computing the second eigenvalue of the orderly part requires dealing only with a representative-agent matrix, which will usually be small.

We note also that $\mathbf{P}$, and hence the result, is robust to a certain types of measurement limitation and/or error.   The interaction matrix $\mathbf{P}$
can be estimated without requiring precise information about agents' actual degrees, but instead their relative proclivity to connect to different types of agents.  For example, it would be enough to 
have a representative sample of each type's neighbors. This makes the model relevant in practical settings, since it is often very difficult to know what fraction of agents' friends are actually reported or observed.

The usefulness of Theorem \ref{thm:representativeAgent} becomes evident through a series of its implications. First, it can be used to tighten the lower bound in Lemma \ref{lem:ConsensusAndEigenvalue}, so that second eigenvalues become an even better proxy for consensus time; this is done in  Proposition \ref{prop:homophilousConvergenceBounds} in the appendix. In the next section, we also use the theorem to derive expressions allowing us to understand how homophily affects consensus and mixing time in the islands model. Lastly, we use it to show that the degree weighted homophily bound in Proposition \ref{prop:eigenvalueBound} is tight.

\subsubsection{Non-Spectral Measures of Homophily}

The relationships between consensus and mixing time relate to the second eigenvalue of the updating matrix, and through our representative agent theorem, to the 
second eigenvalue of the matrix of probabilities of connection across types in the random networks model.    These characterizations are still somewhat abstract as the second eigenvalue
is an implicitly defined statistic that may be hard to grasp.   In order
to understand the implications of the structural feature of homophily on mixing and consensus time, we need to develop some understanding of how homophily affects the second eigenvalue.

We begin with some general definitions of homophily, and later specialize to multi-type random networks.

\paragraph{General Networks and Degree-Weighted Homophily}

Let us partition $N$ into two subsets, $ M $ and $M^c$. First, we define a notion of the weight between two groups.

\begin{definition} 
Given $\mathbf{T}=\mathbf{T(A)}$ and two subsets of nodes, $B, C \subseteq N$, let 
$$ W_{B,C} = \frac{ \sum_{\substack{i \in B \\ j \in C}} T_{ij} T_{ji}}{|B| |C|}.$$ \end{definition}

$W_{B,C}$ keeps track of the relative weight between two sets of nodes $B$ and $C$, and is a measure that ranges between 0 and 1.
The weight of an edge is proportional to the reciprocal of the product of the degrees of the nodes on its ends: when an edge is between two nodes that have many neighbors, it doesn't count for much, but when it is between two that have few neighbors, it counts for a lot. The weight also depends on group size: individual edges within larger groups matter less than those within smaller groups.
With this definition in hand, we define a notion of degree-weighted homophily.

\begin{definition} Given any $\emptyset \subsetneq M \subsetneq N$, 
let the \emph{degree-weighted homophily} of the network $\mathbf{A}$ relative to $M$ be defined by
\begin{equation} \label{eqn:defDWH} \DWH(M; \mathbf{A}) = \frac{W_{M,M}+W_{M^c,M^c}-2W_{M,M^c}}{\frac{1}{|M|^2} \sum_{i \in M} \frac{1}{d_i(\mathbf{A})} + \frac{1}{|M^c|^2} \sum_{i \in M^c} \frac{1}{d_i(\mathbf{A})}}, \end{equation} 
where the $W$'s are relative to $\mathbf{T(A)}$.\end{definition}

The term in the numerator keeps track of how much of the weight in $\mathbf{T}$ falls within $M$ and within $M^c$, and how much weight goes between
these sets of nodes.  So, links within the group $M$ or
its complement $M^c$ increase the degree weighted homophily and links between the two groups decrease it. 
The term in the denominator is a normalizing value which guarantees\footnote{This can be verified by using the expression of DWH as a quadratic form in the proof of Proposition \ref{prop:eigenvalueBound} below and then noting that the spectral norm of the matrix $\mathbf{T}(\mathbf{A})$ is $1$.} that this quantity is always between $-1$ and $1$.

To see that the degree-weighted homophily has an intuitive interpretation, consider a very simple special case. Suppose $|M| = n/2$ and $\mathbf{A}$ corresponds to a regular graph, where all degrees are equal. Then 
\begin{equation} \DWH(M; \mathbf{A}) = \frac{\#(\text{within-group edges}) - \#(\text{between-group edges})}{\#(\text{total edges})}. \label{eqn:regularDWH} \end{equation}

The theoretical justification for the usefulness of this 
measure is that it provides a lower bound on the magnitude of the second eigenvalue of $\mathbf{T}$, and in the limit a tight bound.

Let 
$$\DWH(\mathbf{A})=\max_{\emptyset \subsetneq M \subsetneq N} |\DWH(M; \mathbf{A})|.$$
Thus, the degree weighted homophily of a given network is the maximum level of degree homophily across different possible splits of the network.\footnote{This has intuitive relationships to a weighted version of a min cut, although this degree weighted homophily measure turns out to be the right one for our purposes.}

\begin{proposition}
Assume that $\mathbf{A}$ is connected. Then
\begin{equation}   |\lambda_2(\mathbf{T}(\mathbf{A}))| \geq \displaystyle |\DWH(\mathbf{A})|\label{eqn:eigenvalueBound} \end{equation} \label{prop:eigenvalueBound}  \end{proposition}

Combining this with Lemmas \ref{lem:ConsensusAndEigenvalue} and \ref{mixing} we  see that degree weighted homophily provides a lower bound on the consensus and mixing times.

\begin{corollary} Assume $\mathbf{A}$ is connected. Then for any $0 < \varepsilon < 1$,  
$$ \CT(\varepsilon; \mathbf{A}) \geq \left \lfloor \frac{\log (1/(4\varepsilon)) - \log(1/\underline{s})}{\log (1/ \DWH(\mathbf{A}))} \right \rfloor,$$
and
$$ \MT(\varepsilon; \mathbf{A}) \geq \left \lfloor \frac{\log (1/(2\varepsilon))}{\log (1/ \DWH(\mathbf{A}))} \right \rfloor.$$
\label{cor:DWHTimeBounds} \end{corollary}

Compared with Lemmas \ref{lem:ConsensusAndEigenvalue} and \ref{mixing}, the results have only one inequality each. This is because DWH only provides a lower bound on eigenvalues, and hence on the convergence times. In the next subsection, we supply asymptotic upper bounds in the islands setting.

\paragraph{The Islands Model and Simpler Measures of Homophily}

In order to develop the clearest and most intuitive relationships between homophily and the resulting consensus and mixing times, we now examine a specific 
case of the multi-type random network model.  
Recall the islands model from Section \ref{sec:islands}, and consider the case where there are $m\geq 2$ equally-sized groups. In all the results of this section, we will consider only $n$ divisible by $m$, and the the results will concern limits as $n \to \infty$. All quantities (probabilities, homophilies, etc.) are implicitly indexed by $n$, but we suppress this indexing unless it is important to emphasize it.  
Let $p_s$ and $p_d$ be the probability of links within and across types, respectively, and $p$
be the overall probability of links.

Let $\EDWH(m,p_s,p_d)$ denote the expected degree weighted homophily in the islands model where we calculate
this relative to the {\sl expected} number of links within and across islands.  
That is, if we have a collection of $k$ islands, $M$,
let
$$ EW_{M,M} = \frac{  [p_s k  + p_dk(k-1)]/d^2}{k^2}$$ 
and
$$ EW_{M,M^c} = \frac{p_dk(m-k)/d^2}{k(m-k)}= p_d/d^2,$$
where 
$d= pn$ is the expected degree, and $EW_{MM}$ and $EW_{MM^c}$ are the expected versions of $W_{MM}$ and $W_{MM^c}$.
Then, we have an expected variation of degree weighted homophily:
$$\EDWH(M;m,p_s,p_d) = \frac{EW_{M,M}+EW_{M^c,M^c}-2EW_{M,M^c}}{\frac{1}{|M|^2} \sum_{i \in M} \frac{1}{d} + \frac{1}{|M^c|^2} \sum_{i \in M^c} \frac{1}{d}}. $$
Let $I(n)$ denote the subsets of nodes that are collections of islands, so that if 
$M\in I(n)$ then any node in $M$ is of a different type from any node in $M^c$.
Let
$$\EDWH(m,p_s,p_d)=\max_{M\in I(n)}\EDWH(M;m,p_s,p_d).$$
This is not quite the expected degree weighted homophily, since we are working with expectations in the numerator and denominator.

We can now prove the following theorem which lets us deduce how consensus and mixing time depend on degree weighted homophily in the islands (see Corollary \ref{cor:islandslimits}), since it establishes how the second
eigenvalue relates to homophily and we have already established how consensus and mixing time relate to the second eigenvalue.

\begin{theorem}
\label{thm:islandslimitseig}
In the equal-sized islands model, if $p(n)n/\log^2 (n)\rightarrow\infty$
then 
\begin{equation}\label{eqn:islandsEigenvalue} \left| \lambda_2(\mathbf{T}(\mathbf{A}(\mathbf{P}, \mathbf{n}))) -  \EDWH(m,p_s,p_d) \right| \xrightarrow{p} 0. 
\end{equation}
\end{theorem}

Theorem \ref{thm:islandslimitseig} provides us with limiting expressions for the second eigenvalue as a function of
homophily.  

Let
$$H = \frac{p_s}{p}$$
capture how much more probable a link to own type is compared to other types,
and  
$$h = \frac{p_s}{mp}$$
be the relative fraction of links to own type.
If we index a sequence of societies by
their cardinalities $n$,
then the following lemma establishes the relation between degree weighted homophily and the relative probabilities of links and the number of islands.

\begin{lemma}
\label{lem:edwh}
In the islands model with $m\geq 2$ equal-sized groups and probabilities of links within and across types
$p_s$ and $p_d$, respectively, 
the degree weighted homophily is 
$$\EDWH(m,p_s,p_d) = \frac{p_s-p_d}{p_s+(m-1)p_d}=\frac{H-1}{m-1}.$$
Moreover, $$\EDWH(m,p_s,p_d)= \EDWH(M;m,p_s,p_d)$$ 
for all ${M\in I(n)}$, so the grouping of the islands is irrelevant in calculating the homophily.
If the number of islands
$m(n)$ diverges then 
$|\EDWH(m,p_s,p_d) -   h |\rightarrow 0$.
\end{lemma}

From Theorem
\ref{thm:islandslimitseig} and Lemma \ref{lem:edwh}, the following corollary, characterizing second eigenvalues in terms of traditional measures of homophily in the islands model, follows immediately.

\begin{corollary}
\label{cor:traditionalhomophilies}
In the islands model with $m\geq 2$ equal-sized groups, if $p(n)n/\log^2 (n)\rightarrow\infty$ and probabilities of links within and across types
$p_s$ and $p_d$, respectively, 
\begin{equation}\left| \lambda_2(\mathbf{T}(\mathbf{A}(\mathbf{P}, \mathbf{n}))) -  \frac{H-1}{m-1} \right| \xrightarrow{p} 0. \end{equation}
If the number of islands diverges, then $|\lambda_2(\mathbf{T}(\mathbf{A}(\mathbf{P}, \mathbf{n}))) -   h |\rightarrow 0$.
\end{corollary}

Combining  Lemmas \ref{lem:ConsensusAndEigenvalue} and \ref{mixing} with Corollary \ref{cor:traditionalhomophilies} we
have the following summary of the asymptotic behavior of consensus and mixing time in the islands model, which is Theorem \ref{thm:islandslimits}.

\begin{corollary}
\label{cor:islandslimits}
In the equal-sized islands model, if $p(n)n/\log^2 (n)\rightarrow\infty$ and $h(n)$ is bounded away from 1,
then for any $\delta<1$, with a probability going to 1:
$$
\frac{(1-\delta)\log(n)}{2\log(\frac{m-1}{H-1})}
\leq  \CT(\gamma/n^2; \mathbf{A}(\mathbf{P}, \mathbf{n})) \leq
\frac{(1+\delta)\log(n)}{\log(\frac{m-1}{H-1})}
$$
and 
$$
\frac{(1-\delta)\log(n)}{\log(\frac{m-1}{H-1})}
\leq  \MT(\gamma/n; \mathbf{A}(\mathbf{P}, \mathbf{n})) \leq
\frac{\frac{3}{2}(1+\delta)\log(n)}{\log(\frac{m-1}{H-1})}
.$$
\end{corollary}

The condition that  $h$ is bounded away from 1 rules out the case that all but a vanishing fraction of links are within islands.  If that is
the case, then the islands can become disconnected with a nontrivial probability and the mixing and consensus times diverge.  
The consideration of setting $\varepsilon=1/n^2$ for consensus time and $\varepsilon=1/n$ for mixing time ensure that 
the convergence is within the order of magnitude of the weight on any given node. 
We can rewrite these expressions as
$$
\frac{(1-\delta)\log(n)}{2|\log(\EDWH(m,p_s,p_d))|}
\leq  \CT(\gamma/n^2; \mathbf{A}(\mathbf{P}, \mathbf{n})) \leq
\frac{(1+\delta)\log(n)}{|\log(\EDWH(m,p_s,p_d))|}
$$
and 
$$
\frac{(1-\delta)\log(n)}{|\log(\EDWH(m,p_s,p_d))|}
\leq  \MT(\gamma/n; \mathbf{A}(\mathbf{P}, \mathbf{n})) \leq
\frac{\frac{3}{2}(1+\delta)\log(n)}{|\log(\EDWH(m,p_s,p_d))|}
.$$  

Corollary \ref{cor:islandslimits} provides us with a fairly precise relationship between homophily and 
consensus and mixing times, as further discussed in previous sections.  

\medskip

The above results presume equal sized islands.  We also provide a result for unequal sizes for the case of
two islands. This shows that the degree weighted homophily bound in Proposition \ref{prop:eigenvalueBound} is tight.

\begin{theorem} 
\label{thm:DWHTight} Suppose $\mathbf{n} = (\lceil f_1 n \rceil, \lfloor (1-f_1) n \rfloor)$ with $0< f_1 < 1$ and $$ \mathbf{P} = \left[ \begin{array}{cc}
p_s & p_d  \\
p_d & p_s  \end{array} \right], $$ where all the entries of this matrix are positive. Then $$ {\rm plim}_{n \to \infty}  \lambda_2(\mathbf{T}(\mathbf{A}(\mathbf{P},\mathbf{n}))) = \frac{p_s}{p_s+p_d(f_1^{-1}-1)} - \frac{p_d}{p_s(f_1^{-1}-1)+p_d} = {\rm plim}_{n \to \infty} \DWH(N_1;\mathbf{A}(\mathbf{P},\mathbf{n})),$$ where $N_1$ denotes the first $\lceil f_1 n \rceil$ nodes. 
\end{theorem}

\section{Consensus and Mixing Times in the \\ Adolescent Health Data}

In this section, we examine consensus time and mixing times in 84 social networks from the Adolescent Health dataset\footnote{Add Health is a program project designed by J. Richard Udry, Peter S. Bearman, and Kathleen Mullan Harris, and funded by a grant P01-HD31921 from the National Institute of Child Health and Human Development, with cooperative funding from 17 other agencies. Persons interested in obtaining data files from Add Health should contact Add Health, Carolina Population Center, 123 W. Franklin Street, Chapel Hill, NC 27516-2524 (addhealth@unc.edu).  We thank James Moody for making available the data organized in Pajek files for the 84 schools.} and show how the patterns in that data illustrate our conclusions. For each of the 84 schools, the dataset includes information on each student's grade, gender and race. In addition, each student was asked to name his or her closest male and female friends.\footnote{The number of friends reported was capped at five of each type, or ten in total.  Less than ten percent of the students hit the caps, but that still censors the data.  This design feature makes homophilies computed based on gender somewhat less reliable than the others, since it would tend to equalize the numbers of reported male and female friends, even if there were strong homophily present.}  Using the reported friendship networks (linking two individuals if either named the other as friend) we compute consensus and mixing times. 
We can also examine traditional and degree weighted homophily measures based on the observed characteristics: grade, race and sex.  
Grade is the year in school, and ranges from $6$ to $12$, as most of the schools include 6 different years of students.  Race is self-reported as Asian, black, Hispanic, white, or other (and these were the only categories permitted). Sex is self-reported as male or female.  

It is worth commenting briefly on the nature of the illustration that our empirical computations provide for the theoretical work. 
There are actually three aspects to the full spectrum of our results that can be tested.   
\begin{itemize}
  \item First, there is a question of whether substantial information about
the impact of homophily can be captured by examining relatively simple definitions of types.  
\item Second, many of our results are asymptotic and there is a question about whether our bounds
on how consensus time and mixing time relate to homophily will be useful in finite samples of medium size.   
\item Third, there is a question of whether or not people actually communicate in ways that are captured by the learning model and random Markov models that underlie consensus and mixing time.    
\end{itemize}

Our empirical analysis answers the first two questions in the affirmative.  In particular,
the multi-type random network model is a good fit for these social networks when it comes to investigating consensus and mixing times, and gets a good deal of explanatory power
from very basic definitions of types. Our main claim --- that the study of the convergence of Markovian processes on large networks can be reduced to simple computations about homophily --- is not merely an asymptotic, theoretical claim, but one that holds up well when applied to the data. To see this more clearly, consider some of the typical building blocks that went into establishing the relationship between convergence and homophily: Lemma \ref{lem:ConsensusAndEigenvalue}, Theorem \ref{thm:representativeAgent}, and Corollary \ref{cor:islandslimits}, for example. Each result either provides inequalities or statements about asymptotic convergence. \emph{A priori}, the data might be badly behaved with respect to either of these. It might stay within the inequalities in a noisy way (oscillating randomly between the bounds). It might also take very large networks for the asymptotic results to kick in. 
Lastly, it might even be the case that the multi-type random graph model captures none of the salient structure of these social networks. If any of these happened, then there might be very weak, nonexistent, or ``wrong way'' correlations in the quantities that we studied empirically in this section. The fact that the correlations are quite strong and correspond to our predictions shows that the  relationships suggested by the inequalities and asymptotic theory are relevant. In particular, these high school friendship networks seem to have many of the salient features of well-behaved multi-type random graphs, and much can be captured with simple definitions of types.

Whether or not these models of updating and communication shed light on actual social behavior -- that is, on how people actually communicate in or navigate networks -- is obviously an important question, but one that requires additional (longitudinal) data and is left for future investigations.  

\medskip

The first step is to test the relationship summarized in Theorem \ref{thm:islandslimits} between convergence and homophily in the equal-sized islands model. 
We begin the building blocks of the theorem to rearrange it into an inverted form that compensates for the extreme behavior of the convergence times at high homophilies and makes the quantities amenable to linear regressions. In particular, we define  
$$\rho(X) = \exp\left(-\frac{\log(n)}{X} \right).$$ 
By Lemmas \ref{lem:ConsensusAndEigenvalue} and \ref{mixing}, $\rho(\CT(\gamma/n^2;\mathbf{A}))$
and $\rho(\MT(\gamma/n;\mathbf{A}))$ are approximately the second eigenvalue of $\mathbf{T}(\mathbf{A})$,
which, by
Corollary \ref{cor:traditionalhomophilies}, is well approximated by $\frac{H-1}{m-1}$.
For instance, if $X$ is an empirical measurement of a consensus time for some choice of $\varepsilon = \gamma/n^2$, then $\rho(X)$ can be thought of as an imputed per-step rate of convergence. Thus, we run regressions of $\rho(\CT(\gamma/n^2;\mathbf{A}))$ and $\rho(\MT(\gamma/n;\mathbf{A}))$ on $\frac{H-1}{m-1}$.

The regressions include an intercept term. This is not in Theorem \ref{thm:islandslimits} or its constituent parts. However, it turns out that if there is additional homophily within each island on dimensions not reported in the data, then there will be an intercept term in the model. Details are in Section \ref{sec:affinebias}.

\begin{figure}[htp] \begin{center}
\includegraphics[width=5in]{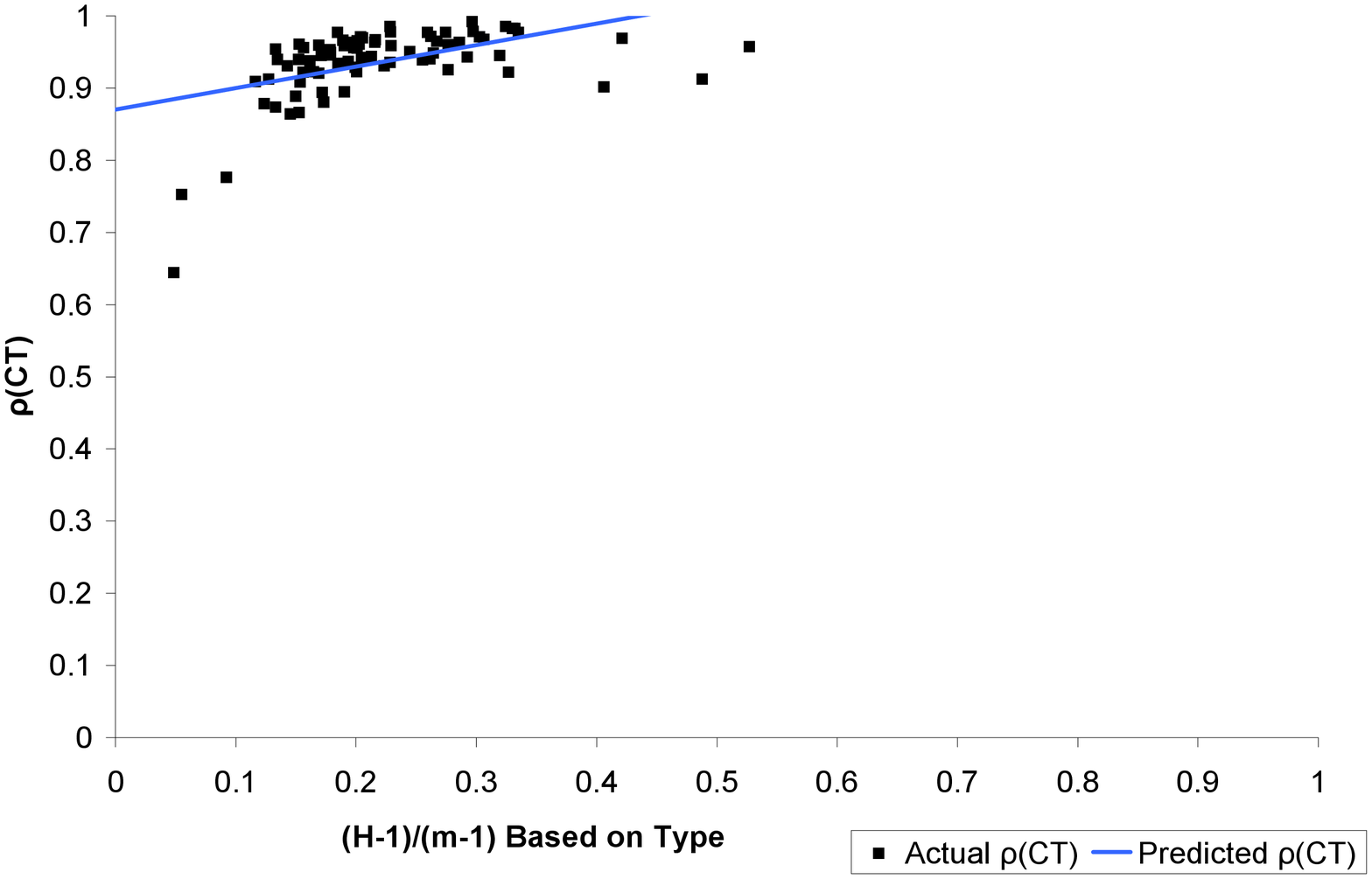}
\caption{The relationship between $\rho(\CT(0.1/n^2;\mathbf{A}))$ and $\frac{H-1}{m-1}$ computed based on the finest-grained type data available (i.e. a type is a race-grade-sex tuple).} 
\label{fig:rhoCTversushomophilytype}
\end{center}
\end{figure}

Lastly, we removed two data points whose consensus and mixing times exceeded our algorithms' capacity. These networks (schools number $53$ and $57$) had very large consensus times (on the order of serveral thousand), so computing them precisely was infeasible. These would not substantially change the results of the first two regressions, since for those purposes, a consensus or mixing time of about $1000$ is essentially infinite. So, from now on, we work with the $82$ data points excluding those schools. The results are presented in Table \ref{tab:rhoCTonHtype} and Figure \ref{fig:rhoCTversushomophilytype}.

\begin{figure}[htp] \begin{center}
\includegraphics[width=5in]{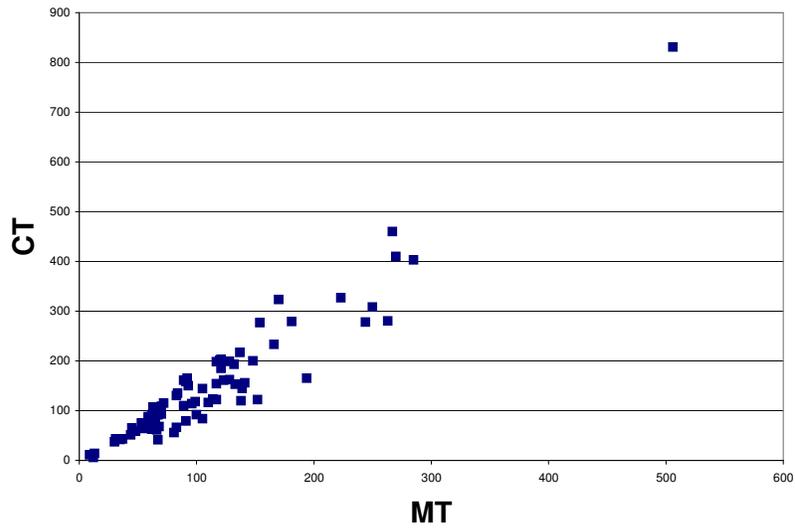}
\caption{The relationship between consensus and mixing times.} 
\label{fig:ctversusmt}
\end{center}
\end{figure}

Results for mixing time are very similar to those for consensus time, and so we collect the analogous results for mixing time in the appendix.  
The fact that consensus time and mixing time show similar results is not surprising, given that they both involve measures of distance between $\mathbf{T(A)}$ and its limit.
To see this directly, we note the tight relationship between the two in Figure \ref{fig:ctversusmt}.

We begin with the finest definition of type available in the data.  Thus, we consider a ``type'' to be a specific combination of race, grade, and sex:  so for instance a type would be all female Asians in grade 9.  Thus, in a high school with two sexes, four races, and four grades there are thirty two types.

{
\def\sep{0.5em}
\def\fns{\footnotesize}
\def\onepc{$^{\ast\ast}$} \def\fivepc{$^{\ast}$}
\def\tenpc{$^{\dag}$}
\def\legend{\multicolumn{3}{l}{\footnotesize{Significance levels
:\hspace{1em} $\dag$ : 10\% \hspace{1em}
$\ast$ : 5\% \hspace{1em} $\ast\ast$ : 1\% \normalsize}}}
\begin{table}[htbp]\centering
 \caption{Dependent variable =  $\rho(\CT(0.1/n^2;\mathbf{A}))$} ($N=82$)\\
\label{tab:rhoCTonHtype}
\begin{tabular}{l r l}\hline\hline 
\textbf{Variable}  & \multicolumn{1}{c}{\textbf{Coefficient}} &\\
  & \multicolumn{1}{c}{\textbf{($t$-statistic)}} &\\\hline
Intercept & 0.870 \\ &\fns{(61.2)}  \\[\sep]
$\frac{H-1}{m-1}$ for ``type'' & 0.297  \\ &\fns{(4.91)} \\[\sep]
\hline R$^{2}$ & {0.231}  &\\
\hline
\hline
\end{tabular}
\end{table}
}

The $R^2$ in the above regression shows that the homophily among these types accounts for roughly a quarter of the variation in consensus and mixing times in the data. This is reasonably high in view of the fact that many qualities that determine network formation -- such as interests, extracurricular activities,  etc. -- are not captured by these data.

\begin{figure}[htp] \begin{center}
\includegraphics[width=5in]{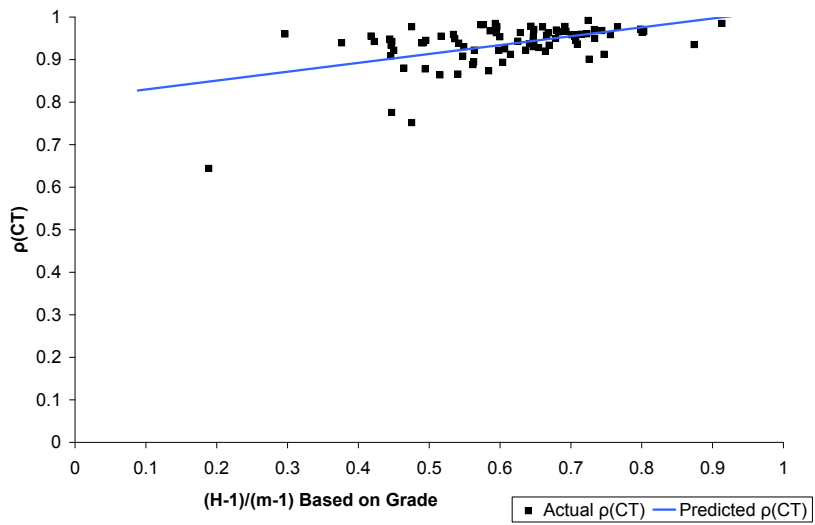}
\caption{The relationship between $\rho(\CT(0.1/n^2;\mathbf{A}))$ and $\frac{H-1}{m-1}$ for grade.} 
\label{fig:rhoCTversushomophily}
\end{center}
\end{figure}

We also explore how much of the variation can be explained by the even simpler definitions of types.
For example, for grades, out of the three characteristics captured in the data, has 
the greatest variation in homophily. The grades also have approximately equal sizes in most of the schools, so that it is legitimate to use the formulas from the equal-sized islands model. 
The results are then reported in Table \ref{tab:rhoCTonH} and Figure \ref{fig:rhoCTversushomophily}.

{
\def\sep{0.5em}
\def\fns{\footnotesize}
\def\onepc{$^{\ast\ast}$} \def\fivepc{$^{\ast}$}
\def\tenpc{$^{\dag}$}
\def\legend{\multicolumn{3}{l}{\footnotesize{Significance levels
:\hspace{1em} $\dag$ : 10\% \hspace{1em}
$\ast$ : 5\% \hspace{1em} $\ast\ast$ : 1\% \normalsize}}}
\begin{table}[htbp]\centering
 \caption{Dependent variable =  $\rho(\CT(0.1/n^2;\mathbf{A}))$} ($N=82$)\\
\label{tab:rhoCTonH}
\begin{tabular}{l r l}\hline\hline 
\textbf{Variable}  & \multicolumn{1}{c}{\textbf{Coefficient}} &\\
  & \multicolumn{1}{c}{\textbf{($t$-statistic)}} &\\\hline
Intercept & 0.809 \\ &\fns{(32.6)}  \\[\sep]
$\frac{H-1}{m-1}$ for grade & 0.209  \\ &\fns{(5.21)} \\[\sep]
\hline R$^{2}$ & {0.253}  &\\
\hline
\hline
\end{tabular}
\end{table}
}

%Lastly, we can directly check the prediction of Theorem \ref{thm:representativeAgent} by computing the representative
%agent matrix based on a certain type 

The fit is similar in quality to that obtained from the finest definitions of type.

We can also examine some other implications of the theory. By focusing on Theorem \ref{thm:islandslimitseig} instead of Theorem \ref{thm:islandslimits} we can replace $\frac{H-1}{m-1}$ in the above regressions by DWH. 
The increase in explanatory power comes from the fact that DWH does not presume equal-sized groups, and thus captures the fact that different islands may be of different sizes.   
We compute DWH based on the three different observed characteristics in the data and use them to estimate $\rho(\CT(\gamma/n^2;\mathbf{A}))$ and $\rho(\MT(\gamma/n;\mathbf{A}))$. For a given definition of type,  we take the DWH over all nontrivial partitions that never separate two agents of the same type. 
For example, $\DWH_{\text{grade}}(\mathbf{A})$ is the DWH taken over all partitions which have some grades on one side and the rest of the grades on the other.  In the tables below, we refer to this quantity as Grade DWH, and similarly with other type classifications.   
More generally,  given some
definition $\theta$ of type such that $\theta : N \to C$ maps agents to types, we define $$ \DWH_{\theta}(\mathbf{A}) := \max_{\substack{\theta^{-1}(M) \cap \theta^{-1}(M^c) = \emptyset \\ \emptyset \subsetneq M \subsetneq N }} \DWH(M; \mathbf{A}).$$  

For this analysis, we do not compute DWH for types, as with thirty or so different groups, the number of different partitions is such that the computations become infeasible.  

Here we are measuring realized degree weighted homophily, DWH, rather than its ``expected'' analogue, EDWH.
This is because EDWH is not available to us, so, as usual, we replace it by the sample analogue. As shown by  the following lemma, this is valid asymptotically.

\begin{lemma}
\label{lem:dwhedwh}
Consider the islands model with $m(n)\geq 2$ equal-sized groups where $m(n)/n\rightarrow 0$, and probabilities of links within and across types
$p_s$ and $p_d$, respectively, and 
consider any sequence of groupings of islands ${M(n)\in I(n)}$.
Then
$$|\EDWH(M(n),m,p_s,p_d) -   \DWH(M(n),\mathbf{A(P,n)}) |\xrightarrow{p} 0,$$
and so
$$|\EDWH(m,p_s,p_d) -   \DWH(M(n),\mathbf{A(P,n)}) |\xrightarrow{p} 0.$$
\end{lemma}

\bigskip

Regressions of convergence rates on the DWH for the 82 networks are reported in Table \ref{tab:rhoCT}. Here, we run the regressions with an intercept term, which is motivated by the same idea as the one formally justifying the inclusion of an intercept in the first regressions of this section. We have not worked out the details formally. Moreover, one of our regressions includes three DWH explanatory variables -- one for each dimension. Such an additively separable form is not justified by the theory but seems to track the data quite closely, so we include it here to point out a potential relationship which may be fruitful to examine further.

\bigskip

\begin{figure}[htp] 
\begin{center}
\includegraphics[width=4in]{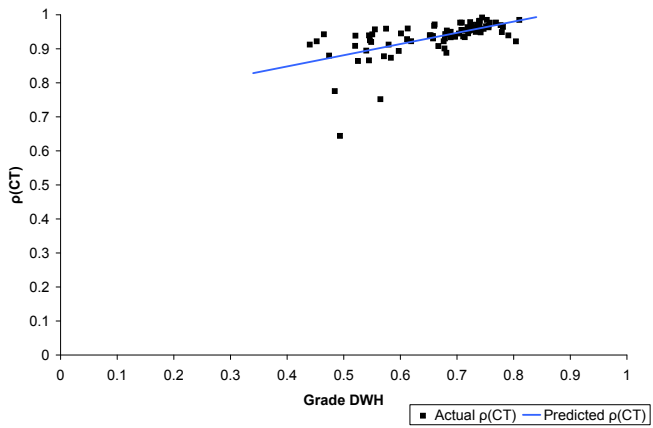}
\caption{Rate of convergence of consensus time for the 82 friendship networks plotted against the degree weighted homophily in each of the networks calculated
relative to grade.} 
\label{fig:rhoCTvsGradeDWH}
\end{center}
\end{figure}

{
\def\sep{0.5em}
\def\fns{\footnotesize}
\def\onepc{$^{\ast\ast}$} \def\fivepc{$^{\ast}$}
\def\tenpc{$^{\dag}$}
\def\legend{\multicolumn{3}{l}{\footnotesize{Significance levels
:\hspace{1em} $\dag$ : 10\% \hspace{1em}
$\ast$ : 5\% \hspace{1em} $\ast\ast$ : 1\% \normalsize}}}
\begin{table}[htbp]\centering
 \caption{Dependent variable =  $\rho(\CT(0.1/n^2;\mathbf{A}))$} ($N=82$)\\
\label{tab:rhoCT}
\begin{tabular}{l r r r r l}\hline\hline 
\textbf{Variable}  & \multicolumn{4}{c}{\textbf{Coefficient}} &\\
  & \multicolumn{4}{c}{\textbf{($t$-statistic)}} &\\\hline
&  \emph{All Homophilies}  & \emph{Grade Only} & \emph{Gender Only} & \emph{Race Only} &\\[\sep]
Intercept & 0.644 & 0.716 & 0.886 & 0.916 \\ &\fns{(20.7)} & \fns{(21.52)} &\fns{(56.4)} & \fns{(88.1)} \\[\sep]
Grade DWH & 0.347 & 0.330 & --- & ---  \\ &\fns{(7.75)} & \fns{(6.64)} &\fns{} & \fns{}\\[\sep]
Gender DWH & 0.137 & ---  & 0.231 & ---\\ & \fns{(2.64)} & \fns{} & \fns{(3.36)} & \fns{} \\[\sep]
Race DWH & 0.105 & ---    & ---   & .0663 \\ & \fns{(5.04)} & \fns{} & \fns{} & \fns{(2.29)} \\[\sep]
\hline R$^{2}$ & {0.545} & {0.356}  & {0.123} & {0.0617} &\\
\hline
\hline
\end{tabular}
\end{table}
}
Consider Table \ref{tab:rhoCT}. The first regression includes all the homophilies acting as independent variables, and the others have each type of homophily used as an explanatory variable on its own. The table shows two main things. First, all three homophilies are significant at the $2\%$ level when the regression is run with all three explanatory variables. Second, grade homophily is doing most of the work in explaining the variation in convergence times; other kinds of homophily have a significant effect, but $36\%$ of the variation can be explained by ignoring all but the grade information. This is illustrated in Figure \ref{fig:rhoCTvsGradeDWH}, where we plot $\rho(\CT(0.1/n^2;\mathbf{A}))$ versus grade homophily and draw the least-squares trend line corresponding to the second column of Table \ref{tab:rhoCT}.

\begin{figure}[htp] \begin{center}
\includegraphics[width=5in]{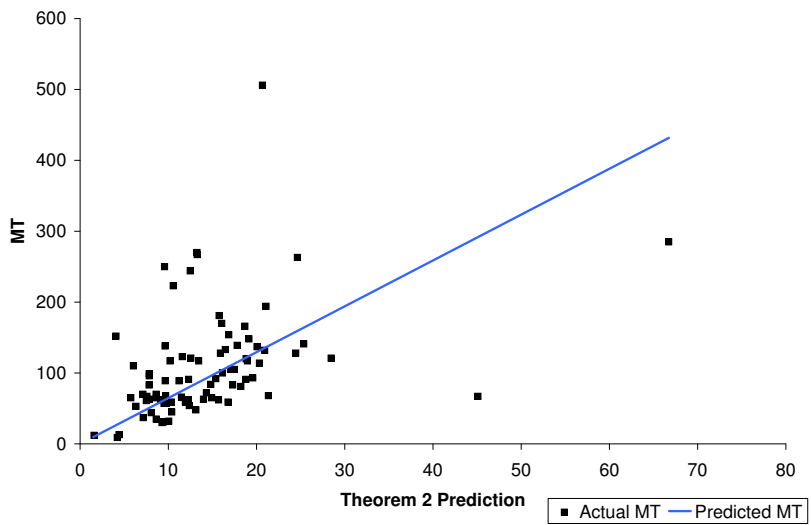}
\caption{The relationship between mixing time and the prediction of Theorem \ref{thm:islandslimits} for grade..} 
\label{fig:ctversusthm2}
\end{center}
\end{figure}

While the above analysis has examined how imputed rates of convergence are affected by homophily, we could work with the consensus times and mixing times directly and compare them to the prediction of Theorem \ref{thm:islandslimits}, which states that they should be approximately proportional to $\log(n)/\log(\frac{m-1}{H-1})$.
When we perform such an analysis, we find results consistent with the theory, as pictured in Figure \ref{fig:ctversusthm2}, where the slope coefficient is significant 
at levels well below $.001$ and the intercept is constrained to be $0$. However, the $R^2$ in this regression
is low ($0.06$) because some extreme data points contribute very large error under this parameterization. It was for this reason that we changed the axes in the above regressions; the rescaling makes errors comparable across data points.

\begin{figure}[htp] \begin{center}
\includegraphics[width=5in]{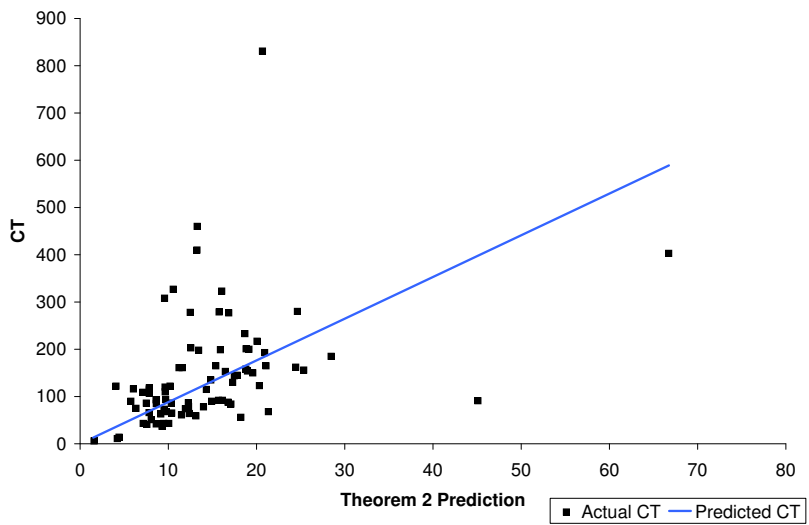}
\caption{The relationship between consensus time and the prediction of Theorem \ref{thm:islandslimits} for grade.} 
\label{fig:mtversusthm2}
\end{center}
\end{figure}

Lastly, we test the prediction of Theorem \ref{thm:shortestpaths} by computing the average shortest path length in each network and running a regression of this on $\log(n)/\log(\bar{d}(\mathbf{A}))$, which is what the theorem predicts the quantity will depend on, as well as on $-\log n / \log\left(\frac{H-1}{m-1} \right)$ (computed based on full-type homophilies) which is supposed to predict consensus and mixing times but not shortest paths. The results are presented in Table \ref{tab:shortestpaths}.

{
\def\sep{0.5em}
\def\fns{\footnotesize}
\def\onepc{$^{\ast\ast}$} \def\fivepc{$^{\ast}$}
\def\tenpc{$^{\dag}$}
\def\legend{\multicolumn{3}{l}{\footnotesize{Significance levels
:\hspace{1em} $\dag$ : 10\% \hspace{1em}
$\ast$ : 5\% \hspace{1em} $\ast\ast$ : 1\% \normalsize}}}
\begin{table}[htbp]\centering
 \caption{Dependent variable =  average shortest path length} ($N=82$)\\
\label{tab:shortestpaths}
\begin{tabular}{l r r l}\hline\hline 
\textbf{Variable}  & \multicolumn{2}{c}{\textbf{Coefficient}} &\\
  & \multicolumn{2}{c}{\textbf{($t$-statistic)}} &\\\hline
&  \emph{Density and Homophily}  & \emph{Density Only} &\\[\sep]
Intercept & $-0.125$ & $-0.106$\\ &\fns{$(-1.08)$}  &\fns{$(-.846)$}\\[\sep]
$\log(n)/\log(\bar{d}(\mathbf{A}))$  & $1.27$  & $1.32$\\ &\fns{$(33.2)$}  &\fns{$(32.9)$} \\[\sep]
$-\log n / \log\left(\frac{H-1}{m-1}\right)$ for type & $0.00981$  \\ &\fns{$(3.79)$} \\[\sep]
\hline R$^{2}$ & {$0.942$}  & {$0.931$} \\\hline
\hline
\end{tabular}
\end{table}
}

\begin{figure}[htp] \begin{center}
\includegraphics[width=5in]{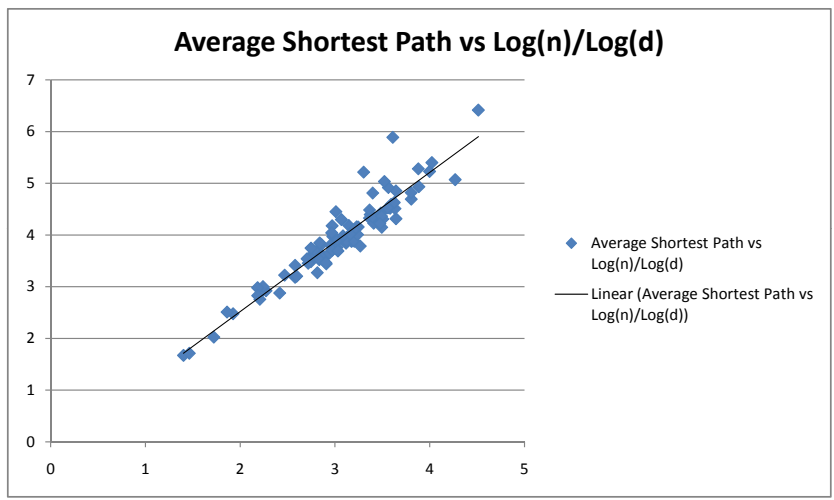}
\caption{The relationship between average shortest path length and the prediction of Theorem \ref{thm:shortestpaths}.} 
\label{fig:shortestpaths}
\end{center}
\end{figure}

The coefficient on the homophily regressor, while significant at conventional levels, has a much smaller $t$-statistic than the coefficient on $\log(n)/\log(\bar{d}(\mathbf{A}))$. Also, since the coefficient on the homophily regressor is about $0.01$ and the values of $-\log n / \log\left(\frac{H-1}{m-1}\right)$ are between $1.58$ and $66.7$ in the data, the overall predictive power of the homophily term is small.  This is also seen in the difference between the two regressions in Table \ref{tab:shortestpaths}, where dropping the homophily dependent variable results in only a one percent change in the R$^2$.  This shows, as predicted, that network density matters much more for shortest path lengths than homophily does. 
The relationship between shortest path and the predicted explantory variable $\log(n)/\log(d)$ is pictured in Figure \ref{fig:shortestpaths}.

\section{Concluding Remarks}

Our results are built in several parts:  
\begin{itemize}
  \item[(i)] we relate communication processes to second eigenvalues largely building on standard spectral theory, 
  \item[(ii)] we provide novel results relating second eigenvalues to homophily, 
  \item[(iii)] we provide novel results relating homophily to random graphs, and 
  \item[(iv)] finally, combining these results enables us to relate communication to homophily in random networks.
\end{itemize}

Our results show that homophily can substantially affect communication processes but that this depends both on the level and type of homophily and the type of communication.  Communication based on shortest paths is essentially unaffected by homophily, while random walks and updating by averaging are affected in a well-identified and nonlinear manner.  
The underlying reason is that homophily does not change shortest paths, but affects the relative numbers of paths between nodes of different types.  Interestingly, there is a complete reversal in the manner in which communication depends on the structure of connections:
\begin{itemize}
  \item  The speed of shortest path communication depends on link density but not homophily.
  \item The speed of Markovian processes (weighted averaging and random walks) depends on homophily but not link density.
\end{itemize}  

The methods used to arrive at these conclusions may be of some independent interest for empirical work. In particular, we have shown that second eigenvalues and convergence times of a stochastic matrix arising from a large multi-type random network can be predicted very accurately from a much smaller matrix that only records relative linking probabilities between types. Thus, instead of attempting to obtain reliable data on an entire large network, which is difficult if not impossible, one is justified in using random sampling to estimate the matrix of probabilities. This approach naturally raises the question of what other global properties of large networks can be estimated accurately using convenient projections of the data which avoid collecting too much local information; this is a potential avenue of further research.

We have also examined a set of 82 networks to see how the communication processes would operate on these networks, and how that relates to the observed homophily of the networks.  
The results show significant relationships that are as predicted by the theory, with increased homophily leading to increased consensus and mixing times according to the predicted formulas.

Our results suggest the importance of understanding homophily in order to understand communication and the functioning of a society.  This is, of course, a first step and suggests many avenues for further research, of which we mention only the most obvious ones.   Considering other sorts of communication, learning, diffusion, and interaction and examining other data
will give a fuller understanding of homophily's role.  For example, an interesting area to explore and to compare results with would be coordination and other games on networks, where it has been found that network structure can affect both the strategic choices (e.g., Morris \citeyearpar{MorrisContagion}, Young \citeyearpar{YoungBook}, Jackson \citeyearpar{JacksonBook}) and the speed of convergence (e.g., Ellison \citeyearpar{Ellison} and Montanari and Saberi \citeyearpar{MontanariSaberi}).

\newpage

\bibliography{homophily-2-7-2009-arxiv-version}

\section{Appendix: Proofs}

\subsection{Background on Reversible Markov Chains}

For completeness and to fix notation, we review very well-known results about Markov chains and self-adjoint matrices which form the foundation for our measures of convergence to consensus and bounds on the time required to converge. None of the material in this section is original; further background and references on these techniques can be found in Diaconis and Stroock \citeyearpar{diaconisstroock}.

Symmetry or self-adjointness is often a useful property to have when working with eigenvalues and other spectral quantities of a matrix. While $\mathbf{T}(\mathbf{A})$ generally will not be symmetric, we can make it into a self-adjoint operator under a well-chosen inner product, as in Diaconis and Stroock \citeyearpar{diaconisstroock}. For this we need a few definitions.

Given a probability distribution $\bm{\pi}$ on $N$, define $$ \langle \mathbf{v} , \mathbf{w} \rangle_{\bm{\pi}} = \sum_i v_i w_i \pi_i.$$ This is just the Euclidean inner product weighted by the entries of the distribution.

\begin{definition} A stochastic matrix $\mathbf{T}$ satisfies \emph{detailed balance} (equivalently, is \emph{reversible}) relative to the distribution $\bm{\pi}$ over the nodes if, for every $i, j \in N$ we have $$\pi_i T_{ij} = \pi_j T_{ji}.$$ \end{definition}

\begin{proposition} If $\mathbf{T}$ satisfies detailed balance relative to $\bm{\pi}$ then $\bm{\pi}$ is a stationary distribution for $\mathbf{T}$. \label{prop:detailedBalanceStationary} \end{proposition}

\noindent{\bf Proof of Proposition \ref{prop:detailedBalanceStationary}}: Observe that $$ \sum_j T_{ji} \pi_j = \sum_j T_{ij} \pi_i = \pi_i \sum_j T_{ij} = \pi_i,$$ where the first equality uses the definition of detailed balance. \eproof

\begin{proposition} \label{prop:detailedBalanceSelfAdjoint} The stochastic matrix $\mathbf{T}$ satisfies detailed balance relative to $\bm{\pi}$ if and only if $\mathbf{T}$ is self-adjoint under the inner product $\langle \cdot, \cdot \rangle_{\bm{\pi}}$. \end{proposition}

\noindent{\bf Proof of Proposition \ref{prop:detailedBalanceSelfAdjoint}}: Assume that detailed balance is satisfied. Take $\mathbf{v} = \bm{\delta}_i$ and $\mathbf{w} = \bm{\delta}_j$ for some $i,j \in N$, i.e. take two standard basis vectors. Then $$\langle \mathbf{T} \mathbf{v}, \mathbf{w} \rangle_{\bm{\pi}} = T_{ji} \pi_j$$ and  $$\langle\mathbf{v},  \mathbf{T} \mathbf{w} \rangle_{\bm{\pi}} = T_{ij} \pi_i.$$ These two quantities are equal by detailed balance and this equality extends to arbitrary $\mathbf{v}$ and $\mathbf{w}$ because the inner product is a bilinear form.

For the converse direction, the equality  $\langle \mathbf{T} \bm{\delta}_i, \bm{\delta}_j \rangle_{\bm{\pi}} = \langle  \bm{\delta}_i, \mathbf{T}\bm{\delta}_j \rangle_{\bm{\pi}}$ for standard basis vectors is guaranteed as a consequence of $\mathbf{T}$ being self-adjoint, and this immediately gives detailed balance by the simple calculation above. \eproof

The next claim is that $\mathbf{T}(\mathbf{A})$ satisfies detailed balance relative to $\mathbf{s}(\mathbf{A})$, which is defined by $$s_i(\mathbf{A}) = \frac{d_i(\mathbf{A})}{\sum_i d_i(\mathbf{A})}.$$  This is immediate to check from the definitions. Thus, $\mathbf{s}(\mathbf{A})$ is the stationary distribution of $\mathbf{T}(\mathbf{A})$ and, moreover, $\mathbf{T}(\mathbf{A})$ is self-adjoint relative to $\langle \cdot, \cdot \rangle_{\mathbf{s}(\mathbf{A})}$. As a result, the eigenvalues of $\mathbf{T}(\mathbf{A})$ are all real. Let $1=\lambda_1(\mathbf{T}(\mathbf{A})),\ldots,\lambda_n(\mathbf{T}(\mathbf{A}))$ denote these eigenvalues ordered from greatest to least by magnitude, and $$1 = \beta_1(\mathbf{T}(\mathbf{A})) > \beta_2(\mathbf{T}(\mathbf{A})) \geq \beta_3(\mathbf{T}(\mathbf{A})) \geq \cdots \geq \beta_n(\mathbf{T}(\mathbf{A})) > -1$$ denote these same eigenvalues ordered from greatest to least as real numbers. Obviously $$|\lambda_2(\mathbf{T}(\mathbf{A}))| = \max \{ |\beta_2(\mathbf{T}(\mathbf{A}))|, |\beta_n(\mathbf{T}(\mathbf{A}))| \}.$$

Now, since $\mathbf{T}(\mathbf{A})$ is self-adjoint, we can use the powerful Courant-Fischer variational characterization of the eigenvalues of a Hermitian matrix\footnote{Horn and Johnson \citeyearpar[p. 176--178]{HornJohnson}.}. Let $\mathbf{e}$ denote the unit column vector of ones.

\begin{proposition} \label{prop:CourantFischer} \begin{equation} \label{eqn:courantFischer1} \beta_n(\mathbf{T}(\mathbf{A})) = \inf_{\mathbf{0} \neq \mathbf{v} \in \R^n} \left\{ \frac{\langle \mathbf{v} , \mathbf{T}(\mathbf{A})\mathbf{v} \rangle}{\langle \mathbf{v},\mathbf{v}\rangle} \right\}. \end{equation} \begin{equation} \label{eqn:courantFischer2} \beta_2(\mathbf{T}(\mathbf{A})) = \sup_{\substack{\mathbf{0} \neq \mathbf{v} \in \R^n \text{ s.t.}\\ \langle \mathbf{v}, \mathbf{e} \rangle=0}} \left\{ \frac{\langle \mathbf{v} , \mathbf{T}(\mathbf{A})\mathbf{v} \rangle}{\langle \mathbf{v},\mathbf{v}\rangle} \right\}, \end{equation} where the inner product everywhere is $\langle \cdot, \cdot \rangle_{\mathbf{s}(\mathbf{A})}$. \end{proposition}  This says that the smallest eigenvalue (under the real number ordering) minimizes the normalized quadratic form in braces, where $\mathbf{v}$ ranges over all nonzero vectors. Moreover, the second largest eigenvalue (under the real number ordering) maximizes the same quantity but where $\mathbf{v}$ ranges over all nonzero vectors orthogonal to the right eigenvector corresponding to the largest eigenvalue.

\subsection{Proofs and Additional Material for the Main Results}

\subsubsection{Relating Consensus Times to Second Eigenvalues}

\noindent{\bf Proof of Lemma \ref{lem:ConsensusAndEigenvalue}}: In the proof, we fix $\mathbf{A}$ and drop it as an argument; we also drop the argument on the eigenvalues, being fixed throughout. 

We first show that  $$  \CT(\varepsilon)  \leq \left \lceil \frac{\log (1/\varepsilon)}{2\log (1/| \lambda_2|)} \right \rceil. $$ Take any $\mathbf{b} \in [0,1]^n$. Let $\mathbf{U}_i$ be the projection onto the eigenspace of $\mathbf{T}$ corresponding to $\lambda_i$. Note that under $\langle \cdot , \cdot \rangle_{\mathbf{s}}$, these eigenspaces are orthogonal. Define $\mathbf{U} = \sum_{i=2}^n \mathbf{U}_i$. This is the projection off the eigenspace corresponding to $\lambda = 1$. Then: 
\begin{align*} \Vert (\mathbf{T}^t - \mathbf{T}^\infty) \mathbf{b} \Vert_{\mathbf{s}}^2 &= \left \Vert \sum_{i=2}^n \lambda_i^t \mathbf{U}_i \mathbf{b} \right \Vert_{\mathbf{s}}^2 && \text{spectral theorem applied to the stochastic matrix} \\ &= \sum_{i=2}^n |\lambda_i|^{2t} \Vert \mathbf{U}_i \mathbf{b} \Vert_{\mathbf{s}}^2 && \text{orthogonality of the spectral projections} \\ &\leq |\lambda_2|^{2t} \sum_{i=2}^n  \Vert \mathbf{U}_i \mathbf{b} \Vert_{\mathbf{s}}^2  \\ &=  |\lambda_2|^{2t} \left \Vert \sum_{i=2}^n  \mathbf{U}_i \mathbf{b} \right \Vert_{\mathbf{s}}^2  && \text{orthogonality of the spectral projections}  \\  &=  |\lambda_2|^{2t} \Vert  \mathbf{U} \mathbf{b} \Vert_{\mathbf{s}}^2  && \text{definition of $\mathbf{U}$}  \\  &\leq |\lambda_2|^{2t} \Vert \mathbf{b} \Vert_{\mathbf{s}}^2   && \text{projections are contractions} \\  &\leq |\lambda_2|^{2t} \sum_{i=1}^n s_i   && \text{$\mathbf{b} \in [0,1]^n$ and definition of $\langle \cdot, \cdot \rangle_{\mathbf{s}}$} \\&= |\lambda_2|^{2t}  && \text{entries of $\mathbf{s}$ sum to $1$.} \end{align*} 
Thus, if $$t \geq \frac{\log (1/\varepsilon)}{2\log (1/| \lambda_2|)}$$ 
then  $$ \Vert (\mathbf{T}^t - \mathbf{T}^\infty) \mathbf{b} \Vert_{\mathbf{s}}^2 \leq \varepsilon, $$ 
from which the bound follows upon observing that $\CT(\varepsilon)$ must be an integer. This also shows that when the second eigenvalue is identically $0$, then consensus time must be $1$.

Now we show that $$\left \lfloor \frac{\log (\underline{s}/4\varepsilon)}{2\log (1/| \lambda_2|)} \right \rfloor \leq \CT(\varepsilon).$$ Let $\mathbf{w}$ be an eigenvector of $\mathbf{T}$ corresponding to $\lambda_2$, scaled so that $\Vert \mathbf{w} \Vert_{\mathbf{s}}^2 = \underline{s}/4$. Then the maximum entry of $\mathbf{w}$ is at most $1/2$ and the minimum entry is at least $-1/2$. Consequently, if we define $\mathbf{b} = \mathbf{w} + \mathbf{e}/2$, then $\mathbf{b} \in [0,1]^n$. Now, using the fact that $\mathbf{e}$ is a right eigenvector corresponding to $\lambda_1=1$ and spectral projections are orthogonal, it follows that: \begin{align*} \Vert (\mathbf{T}^t - \mathbf{T}^\infty) \mathbf{b} \Vert_{\mathbf{s}}^2  &=  |\lambda_2|^{2t} \Vert \mathbf{U}_2 \mathbf{w} \Vert_{\mathbf{s}}^2 \\ &=  |\lambda_2|^{2t} \Vert \mathbf{w} \Vert_{\mathbf{s}}^2 \\ &= \frac{\underline{s}}{4} |\lambda_2|^{2t}. \end{align*} Therefore, if $$t \leq  \frac{\log (\underline{s}/4\varepsilon) }{2\log (1/| \lambda_2|)}$$ then $$ \Vert (\mathbf{T}^t - \mathbf{T}^\infty) \mathbf{b} \Vert_{\mathbf{s}}^2 \geq \varepsilon,$$ from which the remaining bound follows upon observing that $\CT(\varepsilon)$ must be an integer.\eproof

\subsubsection{The Representative Agent Theorem and a Consequence}

 Theorem \ref{thm:representativeAgent} and Proposition \ref{prop:homophilousConvergenceBounds} require related machinery, which we will develop and apply in this section. First, we introduce some notation.

We drop the arguments on the random variable $\mathbf{A}$. Let $\mathbf{D}(\mathbf{A})$ denote the diagonal matrix whose $(i,i)$ entry is $d_i(\mathbf{A})$. Let $\mathbf{R}$ be the $n$-by-$n$ matrix given by $R_{ij} = P_{k\ell}$  if $i \in N_k$, $j \in N_\ell$. Then the expected degree of node $i$ is $w_i := \sum_j R_{ij}$.

 Let $V = \sum_i w_i$ be the sum of expected degrees and $v = \sum_i d_i(\mathbf{A})$ the sum of realized degrees. 
 
 For any matrix $\mathbf{T}$, let $\Vert \mathbf{T} \Vert = \sup_{\Vert \mathbf{v} \Vert = 1} \langle \mathbf{v} , \mathbf{T} \mathbf{v} \rangle,$ where the inner product is the standard Euclidean dot product. Let $$ \mathbf{J} =  \mathbf{D}(\mathbf{A})^{-1/2}\mathbf{A}\mathbf{D}(\mathbf{A})^{-1/2} - v^{-1} \mathbf{D}(\mathbf{A})^{1/2} \mathbf{E} \mathbf{D}(\mathbf{A})^{1/2} $$ and $$ \mathbf{K} =  \mathbf{D}(\mathbf{R})^{-1/2}\mathbf{R}\mathbf{D}(\mathbf{R})^{-1/2} - V^{-1} \mathbf{D}(\mathbf{R})^{1/2} \mathbf{E} \mathbf{D}(\mathbf{R})^{1/2}.$$ 

Now we note a fact from basic linear algebra.

\medskip

\begin{fact} \label{fact:similarity} $ \mathbf{D}(\mathbf{A})^{-1/2}\mathbf{A}\mathbf{D}(\mathbf{A})^{-1/2}$ and $\mathbf{T}(\mathbf{A}) =\mathbf{D}(\mathbf{A})^{-1}\mathbf{A}$ are similar matrices, so that they have the same eigenvalues, and that $v^{-1} \mathbf{D}(\mathbf{A})^{1/2} \mathbf{E} \mathbf{D}(\mathbf{A})^{1/2}$ is the summand of the spectral decomposition of $\mathbf{D}(\mathbf{A})^{-1/2}\mathbf{A}\mathbf{D}(\mathbf{A})^{-1/2}$ corresponding to the eigenvalue $1$. The same reasoning applies when we replace $\mathbf{A}$ by $\mathbf{R}$ and $v$ by $V$. \end{fact}

\medskip

We now state the proof of Theorem \ref{thm:representativeAgent}, or rather a reduction to a proposition which will also be useful for proving Proposition \ref{prop:homophilousConvergenceBounds}.

\medskip

\noindent{\bf Proof of Theorem \ref{thm:representativeAgent}}: It is clear that $\mathbf{D}(\mathbf{R})^{-1}\mathbf{R}$ has the same eigenvalues as $\mathbf{Q}$, so to prove the claim it suffices to prove that the former matrix has second eigenvalue close enough to that of  $\mathbf{D}(\mathbf{A})^{-1/2}\mathbf{A}\mathbf{D}(\mathbf{A})^{-1/2}$.

By Fact \ref{fact:similarity}, we know that $ \Vert \mathbf{J} \Vert$ is the second largest eigenvalue in magnitude of the matrix $ \mathbf{D}(\mathbf{A})^{-1/2}\mathbf{A}\mathbf{D}(\mathbf{A})^{-1/2}$, and $ \Vert \mathbf{K} \Vert$ is the second largest eigenvalue in magnitude of the matrix $ \mathbf{D}(\mathbf{R})^{-1/2}\mathbf{R}\mathbf{D}(\mathbf{R})^{-1/2}$. Thus, by the triangle inequality, if we can show that with probability at least $1- \delta$ we have $\Vert \mathbf{J}- \mathbf{K}  \Vert < \delta$, then the proof is done. This is the content of Proposition  \ref{prop:JK} below. \eproof

Now we state a lemma from the proof of Theorem 3.6 of Chung Lu and Vu \citeyearpar{ChungLuVu}, which is a consequence of a Chernoff-type concentration inequality and is quite useful throughout this section.

\begin{lemma} Fix any $\delta > 0$. If $w_{\min} / \log n$ is high enough, the following statement holds with probability at least $1-\delta$ for all $i$ simultaneously: $|d_i - w_i| < \delta w_i$. \label{lem:degreeDeviation} \end{lemma}

\medskip

\begin{proposition} \label{prop:JK} If $w_{\min}/\log^2 n$ is high enough, then with probability at least $1- \delta$ we have $\Vert \mathbf{J}- \mathbf{K}  \Vert < \delta$. \end{proposition}

\noindent{\bf Proof of Proposition \ref{prop:JK}}: Write \begin{align*} \mathbf{J} - \mathbf{K} = \mathbf{B} + \mathbf{C} + \mathbf{L} + \mathbf{M} \hspace{.3in} \text{where} \hspace{.3in}  
 B_{ij} &= \frac{A_{ij}}{\sqrt{d_i d_j}} \left( 1 - \frac{\sqrt{d_i d_j}}{\sqrt{w_i w_j}} \right) \\ C_{ij} &= \frac{A_{ij} - R_{ij}}{\sqrt{w_i w_j}} \\ L_{ij} &=  \frac{\sqrt{ w_i w_j }}{V} \left( 1 - \frac{\sqrt{ d_i d_j}}{\sqrt{w_i w_j}} \right) \\ M_{ij} &= (V^{-1} - v^{-1} ) \sqrt{ d_i d_j}. \end{align*} 
By the triangle inequality, $$\Vert \mathbf{J} - \mathbf{K} \Vert \leq \Vert \mathbf{B}\Vert + \Vert \mathbf{C}\Vert  + \Vert \mathbf{L} \Vert +  \Vert \mathbf{M} \Vert,$$ so it suffices to bound the pieces individually.

Now we list two lemmas, useful only in this proof, from Chung Lu and Vu \citeyearpar{ChungLuVu}.  The proof of the first requires only minor modification in our setting.

\begin{lemma}  Fix any $\delta > 0$. Then if $w_{\min}/ \log^2 n$ is high enough, with probability at least $1- \delta$: $$ \Vert \mathbf{C} \Vert \leq \frac{2}{\sqrt{\bar{w}}} + \frac{\log n}{\sqrt{w_{\min}}}. $$ \label{lem:CBound} \end{lemma}

\noindent{\bf Proof of Lemma \ref{lem:CBound}}:  The only step of the proof of this last lemma that does not work exactly as in the proofs of Theorems 3.2 and 3.6  of Chung Lu and Vu \citeyearpar{ChungLuVu} is their equation (3.2). This step asserts (in our notation) that for $m \geq 2$, we have $$ \Ex[C_{ij}^m] \leq \frac{(1-R_{ij})R_{ij} + (-R_{ij})^m(1-R_{ij})}{(w_i w_j)^{m/2}} \leq \frac{R_{ij}}{(w_i w_j)^{m/2}} \leq \frac{ w_i w_j/V}{(w_i w_j)^{m/2}} \leq  \frac{ 1/V}{(w_{\min})^{m-2}} .$$ The step which is slightly different is the penultimate inequality. We must show that $R^{ij} \leq w_i w_j/V$. But note that $w_j/V \leq 1$ by definition and $w_i = \sum_k R_{ik} \geq R_{ij}$. \eproof

It follows that we have $\Vert \mathbf{C} \Vert < \delta/4$ with probability at least $1-\delta/4$.

\begin{lemma} Fix any  $\delta > 0$. If $w_{\min}/ \log^2 n$ is high enough, the following statement holds with probability at least $1-\delta$: $$ \Vert \mathbf{M} \Vert \leq \frac{1}{\sqrt{\bar{w}}}.$$ \label{lem:Rbound} \end{lemma}

It follows that we have $\Vert \mathbf{M} \Vert < \delta/4$ with probability at least $1-\delta/4$.

 To bound $\Vert \mathbf{B} \Vert$ and $\Vert \mathbf{L} \Vert$, we will use Lemma \ref{lem:degreeDeviation} and two simple facts about the matrix norm. Let $\abs(\mathbf{X})$ denote the matrix whose $(i,j)$ entry is $|X_{ij}|$.

\begin{lemma} \begin{enumerate} \item For any matrix $\mathbf{X}$, $ \Vert \mathbf{X} \Vert \leq \Vert \abs(\mathbf{X}) \Vert.$ \item  Suppose there are two nonnegative matrices, $\mathbf{X}$ and $\mathbf{Y}$ and a constant $c > 0$ such that for each $i,j$, we have $Y_{ij} < c X_{ij} $. Then $\Vert \mathbf{Y} \Vert \leq c \Vert \mathbf{X} \Vert$.  \end{enumerate} \label{lem:normScaling} \end{lemma}

\noindent{\bf Proof of Lemma \ref{lem:normScaling}}: For (1), note that for all $\Vert \mathbf{v} \Vert = 1$, we have \begin{align*} \langle \mathbf{v}, \mathbf{X} \mathbf{v} \rangle &= \sum_{i,j} v_i v_j X_{ij} \\ &\leq  \sum_{i,j} |v_i v_j X_{ij}| \\ &\leq \sum_{i,j} |v_i| |v_j| |X_{ij}| \\ &\leq \Vert \abs(\mathbf{X}) \Vert, \end{align*} the last inequality being true because $\Vert \abs(\mathbf{v})\Vert = 1$. This proves the claim by definition of the matrix norm.

For (2), note that for all $\Vert \mathbf{v} \Vert = 1$, we have \begin{align*} \langle \mathbf{v} , \mathbf{Y} \mathbf{v} \rangle &= \sum_{i,j} v_i v_j Y_{ij} \\ & <  \sum_{i,j}  |v_i| |v_j| c X_{ij} \\ &\leq  c \sum_{i,j}  v_i v_j  X_{ij}  \\ &\leq c \Vert \mathbf{X} \Vert, \end{align*} where again we have made use of the fact that $\Vert \abs(\mathbf{v})\Vert = 1$. \eproof

To show that, with probability at least $1- \delta$, we have $\Vert \mathbf{B} \Vert < \delta/4$, define $\hat{\mathbf{B}} = \abs(\mathbf{B})$; by Lemma \ref{lem:normScaling}(1) it suffices to show $\Vert \hat{\mathbf{B}} \Vert < \delta/4.$ Note $$\hat{B}_{ij} = \frac{A_{ij}}{\sqrt{d_i d_j}} \left \vert 1 - \frac{\sqrt{d_i d_j}}{\sqrt{w_i w_j}} \right \vert.$$ By Lemma \ref{lem:degreeDeviation} we have with probability at least $1-\delta/4$ that   $$\left \vert 1 - \frac{\sqrt{d_i d_j}}{\sqrt{w_i w_j}} \right \vert < \delta/4$$ and so, noting that $$\Vert \mathbf{D}(\mathbf{A})^{-1/2} \mathbf{A} \mathbf{D}(\mathbf{A})^{-1/2} \Vert = 1$$ and using Lemma \ref{lem:normScaling}(2), the claim is proved.

Precisely the same argument works to show that with probability at least $1-\delta/4$, we have$\Vert \mathbf{L} \Vert < \delta/4$, with $V^{-1} \mathbf{D}(\mathbf{R})^{1/2} \mathbf{E} \mathbf{D}(\mathbf{R})^{1/2}$, which also has norm $1$, playing the role of $\mathbf{D}(\mathbf{A})^{-1/2} \mathbf{A} \mathbf{D}(\mathbf{A})^{-1/2}$.

Combining all the bounds shows that, with probability at least $1-\delta$ we have $\Vert \mathbf{J} - \mathbf{K} \Vert < \delta$, as desired.

This completes the proof of the proposition. \eproof

\medskip
 
We will now use the results established so far in this section to prove a proposition that tightens  the lower bound in Lemma \ref{lem:ConsensusAndEigenvalue}, so that second eigenvalues become an even better proxy for consensus time.

\begin{proposition} \label{prop:homophilousConvergenceBounds} Suppose $(\mathbf{P},\mathbf{n})$ are such that, for all $n$, \begin{enumerate}  \item There exist $\underline{\lambda}$ and $\overline{\lambda}$ so that $0 < \underline{\lambda} \leq \lambda_2(\mathbf{Q}(\mathbf{P},\mathbf{n})) \leq  \overline{\lambda} <1. $
%\item $|\lambda_2(\mathbf{Q}(\mathbf{P},\mathbf{n}))| - |\lambda_3(\mathbf{Q}(\mathbf{P},\mathbf{n}))| \geq \alpha > 0$. 
\item $\min_k n_k/n \geq \alpha > 0$. \item $d_{\min}(\mathbf{P})/d_{\max}(\mathbf{P}) \geq \beta > 0$. \end{enumerate} Write $\mathbf{T} = \mathbf{T}(\mathbf{A})$. Then, for any $\delta > 0$, for high enough $n$, with probability at least $1-\delta$  $$\left \lfloor \frac{\log (1/8\varepsilon) - \log(1/ \alpha \beta)}{2\log (1/| \lambda_2(\mathbf{T})|)} \right \rfloor -1 \leq \CT(\varepsilon; \mathbf{A}) .$$   \end{proposition}

Combining this with Lemma \ref{lem:ConsensusAndEigenvalue}, we can conclude that for any $\delta$ with probability at least  $1-\delta$ 
$$\left \lfloor \frac{\log (1/8\varepsilon) - \log(1/ \alpha \beta)}{2\log (1/| \lambda_2(\mathbf{T})|)}\right\rfloor   -1 \leq 
\CT(\varepsilon; \mathbf{A}) \leq \left \lceil \frac{\log (1/\varepsilon)}{2\log (1/| \lambda_2(\mathbf{T})|)} \right \rceil.$$ 
Thus, as we let $\varepsilon$ get small, we find that $\CT(\varepsilon; \mathbf{A})$ is proportional
to  $ \frac{\log (1/\varepsilon)}{\log (1/| \lambda_2(\mathbf{T})|)}.$

The assumptions of this proposition could be weakened, but in their current form they are simple to state and interpret. The first one says that the second eigenvalue of $\mathbf{Q}(\mathbf{P},\mathbf{n})$ should have a magnitude that stays away from $0$ and $1$, and amounts to requiring that consensus time is not going to $0$ or $\infty$. The other conditions impose some balance on the system. The second one says that no group should be getting negligibly small relative to society. The third one says that maximum and minimum degrees should not get too far apart proportionally. Some types are allowed to be much more popular than others, but not infinitely so. If these conditions are met, then Lemma \ref{lem:ConsensusAndEigenvalue} can be strengthened so that the lower bound is tighter, and still easy to compute.

Techniques similar to the ones used in the proof below can be applied to the study of mixing times in order to tighten the upper bound in Lemma \ref{lem:mixingBounds} in the multi-type random graph setting.

\noindent{\bf Proof of Proposition \ref{prop:homophilousConvergenceBounds}}: We will reuse the same variable names used inside the proof of Proposition \ref{prop:JK}, but  the variables defined for the whole subsection will be unchanged. 

Write $\mathbf{C} = \mathbf{D}(\mathbf{R})^{-1} \mathbf{R}$ and $\mathbf{T} = \mathbf{T}(\mathbf{A})$. That is, $\mathbf{C}$ is the version of $\mathbf{T}$ in the ``expectations'' world. Also, let $$ z = \left \lfloor \frac{\log (1/8\varepsilon) - \log(1/\alpha \beta)}{2\log (1/| \lambda_2(\mathbf{T})|)} \right \rfloor.$$ There are three steps to the proof.  In Step 1, we  show that for $\mathbf{C}^t \mathbf{b}$ to converge within $2 \varepsilon$ of its limit takes at least $z-1$ steps for some $\mathbf{b}$. In Step 2, we use Proposition \ref{prop:JK} to show that for any $\eta > 0$, for high enough $n$, with probability at least $1-\eta$, we have $\Vert \mathbf{T} - \mathbf{C} \Vert <\eta$. In Step 3, we show that, if $\eta$ is chosen small enough, then $\mathbf{C}^t \mathbf{b}$ and $\mathbf{T}^t \mathbf{b}$ are at most $\varepsilon$ apart for at least $z-1$ steps under the inner product $\langle \cdot, \cdot \rangle_{\mathbf{s}(\mathbf{A})}$. This shows the requisite result.

\paragraph{Step 1.} Let $\mathbf{v}$ be a right eigenvector of $\mathbf{C}$ corresponding to eigenvalue $\hat{\lambda}_2 := \lambda_2(\mathbf{Q})$ (this is also the second eigenvalue in magnitude of $\mathbf{C}$ by Fact \ref{fact:similarity}). If we multiply $\mathbf{v}$ by a constant scalar, we may assume that the entry with largest magnitude is $1/2$. By Assumption 1, $\hat{\lambda}_2$ is nonzero. Given this and the fact that $\mathbf{C}$ is constant on a given type, it follows that $\mathbf{v}$ is constant on a given type. Thus, by Assumption 2, there are at least $\alpha n$ entries in $\mathbf{v}$ equal to $1/2$. And from this it follows, by the definition of $\mathbf{s}(\mathbf{C})$ and Assumption 3, that $$\langle \mathbf{v}, \mathbf{v} \rangle_{\mathbf{s}(\mathbf{C})} \geq n \alpha \cdot \left( \frac{1}{2} \right)^2 \cdot \frac{d_{\min}(\mathbf{P})}{n d_{\max}(\mathbf{P})} \geq \frac{\alpha \beta}{4}.  $$ Setting $b_i = v_i + 1/2$, we see as at the end of the proof of Lemma \ref{lem:ConsensusAndEigenvalue} that $$\Vert \mathbf{C}^t \mathbf{b} - \mathbf{C}^\infty \mathbf{b} \Vert_{\mathbf{s}(\mathbf{C})} \geq \frac{\alpha \beta}{4} |\lambda_2(\mathbf{C})|^{2t},$$ which yields the lower bound on convergence time we want with $\mathbf{C}$ instead of $\mathbf{T}$. But in view of Assumption 1 and Theorem \ref{thm:representativeAgent}, for high enough $n$ we can replace $\mathbf{C}$ by $\mathbf{T}$ and lose at most an additive factor of $1$ in the bound.

\medskip

\paragraph{Step 2.} Recall $\mathbf{C} = \mathbf{D}(\mathbf{R})^{-1} \mathbf{R}$ and $\mathbf{T} = \mathbf{T}(\mathbf{A})$. Also put $$\mathbf{L} = \mathbf{D}(\mathbf{A})^{-1/2}\mathbf{J} \mathbf{D}(\mathbf{A})^{1/2}$$ and $$ \mathbf{M} = \mathbf{D}(\mathbf{R})^{-1/2}\mathbf{K}\mathbf{D}(\mathbf{R})^{1/2}.$$ By Fact \ref{fact:similarity}, we have $$\mathbf{T} - \mathbf{C} = v^{-1}  \mathbf{E} \mathbf{D}(\mathbf{A})  - V^{-1}   \mathbf{E} \mathbf{D}(\mathbf{R})  + \mathbf{L} - \mathbf{M}.$$ So by the triangle inequality, it suffices to bound $ \Vert v^{-1}  \mathbf{E} \mathbf{D}(\mathbf{A})  - V^{-1}   \mathbf{E} \mathbf{D}(\mathbf{R}) \Vert$ and $\Vert \mathbf{L} - \mathbf{M} \Vert$. By Lemma \ref{lem:degreeDeviation}, if $w_{\min}/\log^2 n$ is high enough, the following event occurs with probability at least $1-\gamma$ for all $i$ simultaneously: $|d_i - w_i| < \gamma w_i$. Call this event $E_1$. Given the assumptions, high enough $n$ ensures the condition of the lemma is met. Thus, on $E_1$, $$ \Vert v^{-1}  \mathbf{E} \mathbf{D}(\mathbf{A})  - V^{-1}   \mathbf{E} \mathbf{D}(\mathbf{R}) \Vert < \gamma,$$ and so it suffices to take care of the other term.

By Proposition \ref{prop:JK}, we know that if $w_{\min}/\log^2 n$ is high enough, then on an event $E_2$ of probability at least $1- \gamma$ we have $\Vert \mathbf{J}- \mathbf{K}  \Vert < \gamma$. As above, for high enough $n$ the condition is met. Now let $$\mathbf{F} = \mathbf{D}(\mathbf{A}) - \mathbf{D}(\mathbf{R}),$$  $$\mathbf{G} = (\mathbf{D}(\mathbf{R})+\mathbf{F})^{1/2}- \mathbf{D}(\mathbf{R})^{1/2},$$ and $$\mathbf{H} = (\mathbf{D}(\mathbf{R})+\mathbf{F})^{-1/2}- \mathbf{D}(\mathbf{R})^{-1/2}.$$ Observe \begin{align*} \Vert  \mathbf{L} - \mathbf{M} \Vert &= \Vert ( \mathbf{D}(\mathbf{R})+\mathbf{F})^{-1/2}\mathbf{J}(\mathbf{D}(\mathbf{R})+\mathbf{F})^{1/2}- \mathbf{D}(\mathbf{R})^{-1/2}\mathbf{K} \mathbf{D}(\mathbf{R})^{1/2} \Vert \\ &= \Vert ( \mathbf{D}(\mathbf{R})^{-1/2} + \mathbf{H})\mathbf{J}(\mathbf{D}(\mathbf{R})^{1/2}+\mathbf{G})- \mathbf{D}(\mathbf{R})^{-1/2}\mathbf{K} \mathbf{D}(\mathbf{R})^{1/2} \Vert \\ &= \Vert  \mathbf{D}(\mathbf{R})^{-1/2}(\mathbf{J} - \mathbf{K})\mathbf{D}(\mathbf{R})^{1/2} +  \mathbf{D}(\mathbf{R})^{-1/2} \mathbf{J} \mathbf{G} + \mathbf{H} \mathbf{J} \mathbf{D}(\mathbf{R})^{1/2}\Vert \\ &\leq \Vert  \mathbf{D}(\mathbf{R})^{-1/2}(\mathbf{J} - \mathbf{K})\mathbf{D}(\mathbf{R})^{1/2} \Vert + \Vert \mathbf{D}(\mathbf{R})^{-1/2} \mathbf{J} \mathbf{G} \Vert + \Vert \mathbf{H} \mathbf{J} \mathbf{D}(\mathbf{R})^{1/2}\Vert +  \Vert \mathbf{H} \mathbf{J} \mathbf{G}\Vert \end{align*}  Using Lemma \ref{lem:degreeDeviation} and standard series approximation arguments, for high enough $n$ we can ensure $ \Vert \mathbf{G} \Vert \leq \gamma \Vert \mathbf{D}(\mathbf{R})^{1/2} \Vert$ and $ \Vert \mathbf{H} \Vert \leq \gamma \Vert \mathbf{D}(\mathbf{R})^{-1/2} \Vert$ on an event $E_3$ of probability at least $1-\gamma$. Using the fact that  $\Vert \mathbf{J} \Vert \leq 1$, the Cauchy-Schwartz inequality yields that each of the middle two terms above is bounded by $\gamma$. For the last term, note that $$\Vert  \mathbf{H} \mathbf{J} \mathbf{G}\Vert \leq \gamma^2 \Vert \mathbf{D}(\mathbf{R})^{1/2} \Vert \cdot \Vert \mathbf{D}(\mathbf{R})^{-1/2} \Vert = \frac{d_{\max}(\mathbf{P})}{d_{\min}(\mathbf{P})} \leq \frac{\gamma^2}{\beta}.$$

 So it suffices to take care of the first term. This is accomplished by noticing that, on $E_1 \cap E_2$, \begin{align*} \Vert  \mathbf{D}(\mathbf{R})^{-1/2}(\mathbf{J} - \mathbf{K})\mathbf{D}(\mathbf{R})^{1/2} \Vert &\leq \frac{d_{\max}(\mathbf{A})^{1/2}}{d_{\min}(\mathbf{A})^{1/2}}\Vert  \mathbf{J} - \mathbf{K} \Vert \\ &\leq (1+\gamma)\frac{d_{\max}(\mathbf{P})^{1/2}}{d_{\min}(\mathbf{P})^{1/2}}\Vert  \mathbf{J} - \mathbf{K} \Vert && \text{definition of $E_1$} \\ &\leq \frac{1+\gamma}{\beta}\Vert  \mathbf{J} - \mathbf{K} \Vert  && \text{Assumption 3} \\ &\leq \frac{(1+\gamma)\gamma}{\beta} && \text{definition of $E_2$}. \end{align*} Together, these facts show that for high enough $n$, on $E_1 \cap E_2 \cap E_3$, which occurs with probability at least $1-3 \gamma$, we have $$ \Vert \mathbf{T} - \mathbf{C} \Vert \leq \gamma + \frac{(1+\gamma)\gamma}{\beta}  + 2 \gamma + \frac{\gamma^2}{\beta}.$$ By choosing $\gamma$ so that the right hand side is less than $\eta$ and $3 \gamma < \eta$ (to take care of the probability), the step is complete.

\medskip
\paragraph{Step 3.} Write $ \mathbf{T} = \mathbf{C} + \mathbf{Y},$ where $\Vert \mathbf{Y} \Vert \leq \eta.$ Note that $$ (\mathbf{T} + \mathbf{Y})^t = \mathbf{T}^t + \sum_{q=0}^{t-1} \mathbf{X}_q,$$ where $\mathbf{X}_{q}$ is a product of $q$ copies of $\mathbf{Y}$ and $t-q$ copies of $\mathbf{T}$ in some order. By the fact that $\Vert \mathbf{T} \Vert =1$ and $\Vert \mathbf{Y} \Vert \leq \eta$, we have $\Vert \mathbf{X}_{q} \Vert \leq \eta^q$ for each $q \geq 1$. Then, by the triangle inequality, $$ \left \Vert \sum_{q=0}^{t-1} \mathbf{X}_q \right \Vert \leq \sum_{q=0}^{t-1} \eta^q \leq \frac{1}{1 - \eta}.$$ Thus, $$\mathbf{Y}_t := \Vert \mathbf{C}^t - \mathbf{T}^t \Vert \leq \frac{1}{1 - \eta}. $$

Take $\mathbf{b}$ and $\mathbf{v}$ to be the vectors constructed in Step 1. Note that for $t \leq z-1$ we have, for high enough $n$, \begin{align*} \langle \mathbf{T}^t \mathbf{b} -  \mathbf{T}^\infty \mathbf{b}, \mathbf{T}^t \mathbf{b} - \mathbf{T}^\infty \mathbf{b} \rangle_{\mathbf{s}(\mathbf{A})} &= \langle \mathbf{T}^t \mathbf{v}, \mathbf{T}^t \mathbf{v} \rangle_{\mathbf{s}(\mathbf{A})}  \\ &=  \langle (\mathbf{C}^t + \mathbf{Y}_t) \mathbf{v}, (\mathbf{C}^t + \mathbf{Y}_t) \mathbf{v} \rangle_{\mathbf{s}(\mathbf{A})} \\&\geq \langle \mathbf{C}^t  \mathbf{v}, \mathbf{C}^t  \mathbf{v}\rangle_{\mathbf{s}(\mathbf{A})} + 2 \langle \mathbf{Y}_t  \mathbf{v}, \mathbf{C}^t  \mathbf{v}\rangle_{\mathbf{s}(\mathbf{A})} \\&\geq (1-\eta)\langle \mathbf{C}^t  \mathbf{v}, \mathbf{C}^t  \mathbf{v}\rangle_{\mathbf{s}(\mathbf{C})} \\ & \; \; \; \; \;+ 2 \langle \mathbf{Y}_t  \mathbf{v}, \mathbf{C}^t  \mathbf{v}\rangle_{\mathbf{s}(\mathbf{A})} && \text{Lemma \ref{lem:degreeDeviation}}\\ &\leq 2(1-\eta)\varepsilon + 2 \langle \mathbf{Y}_t  \mathbf{v}, \mathbf{C}^t  \mathbf{v}\rangle_{\mathbf{s}(\mathbf{A})}  && \text{definition of $z$ } \\ &\leq 2(1-\eta)\varepsilon - 2\Vert \mathbf{Y}_t   \mathbf{v} \Vert_{\mathbf{s}(\mathbf{A})} \cdot \Vert \mathbf{C}^t   \mathbf{v} \Vert_{\mathbf{s}(\mathbf{A})} && \text{Cauchy-Schwartz}   \\ &\leq 2(1-\eta)\varepsilon - 2\Vert \mathbf{Y}_t   \mathbf{v} \Vert_{\mathbf{s}(\mathbf{A})} && \text{see below} \\ &\leq 2(1-\eta)\varepsilon -  2 \Vert \mathbf{Y}_t   \mathbf{v} \Vert && \text{def'n of $ \Vert \cdot \Vert_{\mathbf{s}(\mathbf{A})}$}  \\ &\leq 2(1-\eta)\varepsilon -  2\eta.  \end{align*}  The step whose explanation is missing is straightforward; no entries in $\mathbf{v}$ have magnitude exceeding $1/2$ and multiplication by the stochastic matrix $\mathbf{C}$ preserves this property. Since $\mathbf{s}(\mathbf{C})$ is a probability distribution, $\Vert \mathbf{C}^t   \mathbf{v} \Vert_{\mathbf{s}(\mathbf{C})} \leq 1$ holds by definition of the inner product. If $\eta$ is chosen so that $2(1-\eta)\varepsilon-2\eta > \varepsilon$, then the proof is complete. \eproof

\subsubsection{Results on DWH and EDWH}

\noindent{\bf Proof of Proposition \ref{prop:eigenvalueBound}}:  We will construct a $\mathbf{v}$ satisfying $\langle \mathbf{v},\mathbf{e} \rangle_{\mathbf{s}} = 0$ so that the absolute value of the quantity  $\langle \mathbf{v} , \mathbf{T}\mathbf{v} \rangle_{\mathbf{s}}/\langle \mathbf{v},\mathbf{v}\rangle_{\mathbf{s}}$ is equal to $|\DWH(M)|$. Since $|\lambda_2| = \max\{|\beta_2|,|\beta_n|\}$, this suffices by Proposition \ref{prop:CourantFischer}.

Define $$ v_i = \begin{cases} \frac{1}{r d_i} & \text{if } i \in M \\ -\frac{1}{(n-r)d_i} & \text{if } i \notin M. \end{cases}$$  Let $D = \sum_i d_i$ and note \begin{align} \nonumber \langle \mathbf{v}, \mathbf{v} \rangle_{\mathbf{s}} &= \sum_i v_i^2 \cdot s_i \\ \nonumber &= \sum_{i \in M} \frac{1}{(r d_i)^2} \cdot \frac{d_i}{D} + \sum_{i \in M^c} \frac{1}{((n-r) d_i)^2} \cdot \frac{d_i}{D} \\ \label{eqn:normOfV} &= \frac{1}{D} \left[ \frac{1}{r^2} \sum_{i \in M} \frac{1}{d_i} + \frac{1}{(n-r)^2} \sum_{i \in M^c} \frac{1}{d_i} \right]. \end{align} Also, \begin{align*} \langle \mathbf{v}, \mathbf{T} \mathbf{v} \rangle_{\mathbf{s}} &= \sum_{i} v_i \left( \sum_j T_{ij} v_j \right) s_i \\ &= \frac{1}{D} \sum_{i} \sum_j v_i  T_{ij} v_j d_i  \\ &= \frac{1}{D} \sum_{i,j : T_{ij} >0} v_i  \frac{1}{d_i} v_j d_i  \\ &= \frac{1}{D} \left[ \frac{1}{r^2} \sum_{i,j \in M} T_{ij}T_{ji} + \frac{1}{(n-r)^2} \sum_{i,j \in M^c} T_{ij}T_{ji}  - \frac{2}{r(n-r)} \sum_{i \in M, j\in M^c} T_{ij}T_{ji} \right]. \end{align*} 
Dividing $\langle \mathbf{v}, \mathbf{T} \mathbf{v} \rangle_{\mathbf{s}}$ by $\langle \mathbf{v}, \mathbf{v} \rangle_{\mathbf{s}}$, canceling $D$, and using the definition of $W$ yields the result. \eproof

We prove Lemma \ref{lem:edwh} before Theorem \ref{thm:islandslimits}.

\noindent{\bf Proof of Lemma \ref{lem:edwh}}: 
We show that $ \EDWH(m,p_s,p_d) = (p_s-p_d)/mp = (p_s-p_d)/(p_s+(m-1)p_d)$, which is easily checked to be equal to $(H-1)/(m-1)$; this, in turn, converges to $h=H/m$ as $m$ grows.
Consider $M$ consisting of $k$ islands and $M^c$ consisting of $m-k$ islands of nodes.
Then from the definition of $\EDWH(M; m, p_s,p_d)$ it follows that
$$\EDWH(M; m, p_s,p_d) = \frac{EW_{M,M}+EW_{M^c,M^c}-2EW_{M,M^c}}{\frac{1}{|M|^2} \sum_{i \in M} \frac{1}{d} + \frac{1}{|M^c|^2} \sum_{i \in M^c} \frac{1}{d}},$$
which can be written as
$$\frac{\frac{k p_s + k (k-1) p_d}{k^2 d^2} +\frac{(m-k) p_s + (m-k) (m-k-1) p_d}{(m-k)^2 d^2}-2\frac{p_d}{d^2}}{\frac{m^2}{k^2 n^2} \frac{kn}{dm} + \frac{m^2}{(m-k)^2 n^2}  \frac{(m-k)n}{dm}}.$$
This becomes
$$\frac{\frac{p_s +  (k-1) p_d}{k} +\frac{p_s +  (m-k-1) p_d}{m-k}-2{p_d}}{\frac{dm}{kn } + \frac{dm}{(m-k)n}},$$
or 
$$\frac{p_s - p_d}{pm},$$
which is the claimed expression.  Since this holds for all $M\in I(n)$, the result follows.\eproof

\noindent{\bf Proof of Theorem \ref{thm:islandslimits}}:  Note $$ \mathbf{Q}(\mathbf{P},\mathbf{n}) = \frac{p_d}{p_s + (m-1)p_d} \mathbf{E}_{m} + \frac{p_s-p_d}{p_s+(m-1)p_d} \mathbf{I}_{m},$$ where $\mathbf{E}_{m}$ denotes the $m$-by-$m$ matrix of ones and $\mathbf{I}_m$ denotes the $m$-by-$m$ identity matrix. Then, the eigenvalues of this matrix can be computed directly. The only nonzero eigenvalue of the first matrix is $$ \frac{m p_d}{p_s + (m-1)p_d} $$ with multiplicity 1; and adding $$ \frac{p_s-p_d}{p_s+(m-1)p_d} \mathbf{I}_m $$ just shifts all the eigenvalues by adding to them the constant multiplying the identity. Thus the second largest eigenvalue of $\mathbf{Q}(\mathbf{P},\mathbf{n})$ (after the eigenvalue $1$) is $$ \frac{p_s-p_d}{p_s+(m-1)p_d}.$$ Simple algebra shows that this is the same as the expression claimed in the theorem.\eproof

\bigskip

\noindent{\bf Proof of Corollary \ref{cor:islandslimits}}:  
Let us first show that with a probability going to 1 all nodes have degree between
$(1-f(n))d$ and $(1+f(n))d$ for a function $f(n)\rightarrow 0$.
From Lemma 2.1 in Chung and Lu (2002),\footnote{Set the $X_i$'s in their lemma to be the realization of the links that a given
node might have to other nodes.} it follows that for any given $i$, $\Pr(d_i\geq (1-f(n))d)> 1-e^{-(f(n))^2d/3}$ for any function $f(n)<1$.  The probability 
that all nodes have degrees at least $(1-f(n))d$ is then at least $\left(1-e^{-(f(n))^2d/3}\right)^n$.  Given that $d\geq \log^2(n)$, it follows that this expression is
at least $\left(1-\frac{e^{-(f(n))^2 \log(n)/3}}{n}\right)^n$, which goes to 1 as long as $(f(n))^2\log(n)/ 3$  goes to $\infty$.  
A similar argument establishes that  all nodes have degrees at most $(1+f(n))d$ with a probability going to 1.  
Thus, with a probability going to 1, $s_{\min} \geq \frac{1-f(n)}{(1+f(n))n}$.

Note, also that with a probability going to 1 that $\mathbf{A}$ is connected (e.g., apply Theorem \ref{thm:avgdist} noting that $h(n)$ is bounded away from 1 so that (iv) applies, and (i)-(iii) apply given the islands model and $d\geq \log^2(n)$).  Thus, we can also apply 
Lemmas \ref{lem:ConsensusAndEigenvalue} and \ref{mixing}, to conclude that with a probability going to 1
$$
\left \lfloor \frac{\log (n^2/4\gamma) - \log((1+f(n))n/(1-f(n)))}{-2\log (| \lambda_2(\mathbf{T}(\mathbf{A}(\mathbf{P}, \mathbf{n})))|)} \right \rfloor \leq \CT(\gamma/n^2; \mathbf{A}) \leq \left \lceil \frac{\log (n^2/\gamma)}{-2\log (| \lambda_2(\mathbf{T}(\mathbf{A}(\mathbf{P}, \mathbf{n})))|)} \right \rceil,
$$
and
$$
\frac{\log(\frac{n}{2\gamma})}{-\log(|\lambda_2(\mathbf{T}(\mathbf{A}(\mathbf{P}, \mathbf{n})))|)}
\leq  \MT(\gamma/n; \mathbf{A}(\mathbf{P}, \mathbf{n})) \leq
\frac{\log(\frac{n}{2\gamma})+ \log((1+f(n))n/(1-f(n)))/2}{-\log(|\lambda_2(\mathbf{T}(\mathbf{A}(\mathbf{P}, \mathbf{n})))|)}.
$$
These imply that with a probability going to 1:
\begin{align}\label{obamawins}
\left \lfloor \frac{\log (n) -\log(4\gamma) - \log((1+f(n))/(1-f(n)))}{-2\log (| \lambda_2(\mathbf{T}(\mathbf{A}(\mathbf{P}, \mathbf{n})))|)} \right \rfloor &\leq \CT(\gamma/n^2; \mathbf{A}) 
 \\ &\leq \left \lceil \frac{\log (n)-\log(\gamma)/2}{-\log (| \lambda_2(\mathbf{T}(\mathbf{A}(\mathbf{P}, \mathbf{n})))|)} \right \rceil, \notag
\end{align} 
and
\begin{align}\label{yippee}
\frac{\log(n) - \log(2\gamma))}{-\log(|\lambda_2(\mathbf{T}(\mathbf{A}(\mathbf{P}, \mathbf{n})))|)}
&\leq  \MT(\gamma/n; \mathbf{A}(\mathbf{P}, \mathbf{n}))\\  &\leq
\frac{\frac{3}{2}\log(n) -\log(2\gamma)+ \log((1+f(n))/(1-f(n)))/2}{-\log(|\lambda_2(\mathbf{T}(\mathbf{A}(\mathbf{P}, \mathbf{n})))|)}.\notag
\end{align}

Next, applying Theorem \ref{thm:islandslimits} and Lemma \ref{lem:edwh},
\begin{equation}\label{second} \left| \lambda_2(\mathbf{T}(\mathbf{A}(\mathbf{P}, \mathbf{n}))) -  \frac{H(n)-1}{m(n)-1} \right| \xrightarrow{p} 0 .
\end{equation}
Since, $H(n)=h(n)m(n)$ it follows that $\frac{H(n)-1}{m(n)-1}= \frac{h(n)m(n)-1}{m(n)-1}$ is bounded away from 1.
Thus, from (\ref{second}), we deduce that for any $1>\delta>0$, with a probability going to 1
$$  \frac{1-\delta}{-\log(\frac{H(n)-1}{m(n)-1})} \leq \frac{1}{-\log(|\lambda_2(\mathbf{T}(\mathbf{A}(\mathbf{P}, \mathbf{n})))|)} \leq \frac{1+\delta}{-\log(\frac{H(n)-1}{m(n)-1})}.$$ 
The corollary then follows from 
(\ref{obamawins}) and (\ref{yippee}), noting that $f(n)\rightarrow 0$.\eproof

\noindent{\bf Proof of Theorem \ref{thm:DWHTight}}:  For the left hand side, apply Theorem \ref{thm:representativeAgent} and then compute the second eigenvalue of $$\mathbf{Q}(\mathbf{P}) = \left[ \begin{array}{cc}
\frac{f_1 p_s}{f_1 p_s + f_2 p_d} & \frac{f_2 p_d}{f_1 p_s + f_2 p_d}  \\
\frac{f_1 p_d}{f_1 p_d + f_2 p_s} &\frac{f_2 p_s}{f_1 p_d + f_2 p_s}  \end{array} \right], $$ (the result appears in Jackson (2008), Section 8.3.6). For the right hand side, first use the definition of DWH; then apply Lemma  \ref{lem:degreeDeviation} to show the degrees in the denominators in the DWH formula are arbitrarily close to their expectations; then use the strong law of large numbers to conclude that the ratios appearing in the formula converge to their expectations.\eproof

\subsection{Results and Proofs for the Empirical Analysis}

\label{sec:appendixEmpirical}

\subsubsection{Misidentifying Islands and Affine Bias}

\label{sec:affinebias}

To justify including an intercept in our regressions, consider the following stylized elaboration of the islands model. We have $m$ equally sized islands $N_1,\ldots,N_m$ and each of those islands $k$ is divided into $r$ equally sized sub-islands $N_{k1},\ldots,N_{kr}$.  If $i$ and $j$ are in different islands, then the linking probability between them is $p_d$. If $i$ and $j$ are in the same island, then the linking probability depends on whether they are in the same sub-island, or different sub-islands. If they are in the same sub-island, then they are linked with probability $p_s$. And if they are in the same island but different sub-islands, then they are linked with probability $p_b$. We assume $0 < p_d \leq p_b \leq p_s$. 

The idea is that the researcher has data on the islands but not the sub-islands. We will now study, in this simple setting, what happens if the homophily $H$ is estimated as if the data were generated by the islands model with islands $N_1,\ldots,N_m$ and no sub-islands.

If, without knowing about the sub-islands, we estimate the probability of same-type nodes being linked, we are actually estimating the quantity $$ \tilde{p}_s = \frac{p_s + (r-1) p_b}{r},$$ and our estimate of $H$, the unnormalized homophily, will be $$ \tilde{H} = \frac{\tilde{p}_s}{mp},$$ where $p$ will be estimated correctly by its sample analogue of link density. 

In this setting, it is not valid to apply Corollary \ref{cor:traditionalhomophilies} with the predictor of the second eigenvalue computed based on the misidentified island structure. That is, the second eigenvalue will not be well estimated by $\frac{\tilde{H}-1}{m-1}$. Instead, the second eigenvalue of the representative agent matrix\footnote{This, by Theorem \ref{thm:representativeAgent} is the limit of the second eigenvalue of the realized matrix.} will be an affine function of $\frac{\tilde{H}-1}{m-1}$. This is the content of the following proposition.

\begin{proposition} \label{prop:affinebias} In the modified islands setting just described, if $\tilde{x} = \frac{\tilde{H}-1}{m-1}$ is the regressor computed without information about the sub-island structure, then $$ \lambda_2( \mathbf{Q}(\mathbf{P},\mathbf{n})) =   \alpha \tilde{x} + \beta,$$ where $\alpha$ and $\beta$ depend on $m$ and $r$. \end{proposition}

\noindent{\bf Proof of Proposition \ref{prop:affinebias}}: Letting $\mathbf{E}_k$ denote the matrix of all ones of size $k$ and $\mathbf{I}_k$ denote the identity matrix of size $k$, we find that with $\mathbf{P}$ specified by the description above, $$ \mathbf{Q}(\mathbf{P},\mathbf{n}) = \frac{ p_d  \mathbf{E}_{mr} + (p_b - p_d)  \mathbf{I}_m \otimes \mathbf{E}_r +( p_s - (p_b - p_d))  \mathbf{I}_{mr}}{p_s + (r-1) p_b + (m-1)r p_d} . $$ Now, the second eigenvalue of $$ p_d  \mathbf{E}_{mr} + (p_b - p_d)  \mathbf{I}_m \otimes \mathbf{E}_r$$ is $ r(p_b - p_d).$ Thus, $$ \lambda_2( \mathbf{Q}(\mathbf{P},\mathbf{n})) = \frac{r(p_b - p_d) +  p_s - (p_b - p_d) }{p_s + (r-1) p_b + (m-1)r p_d}.$$ This can be rewritten as $$\lambda_2( \mathbf{Q}(\mathbf{P},\mathbf{n})) = \frac{  \tilde{H} m ( m r -1) + 1 - r}{(m-1) r}.$$ Letting $\tilde{x} = \frac{\tilde{H}-1}{m-1}$  be the regressor, we find that $$ \lambda_2( \mathbf{Q}(\mathbf{P},\mathbf{n})) =   \frac{ m (m r-1)}{r} \cdot \tilde{x} + \frac{ m r + r - 1}{r}. $$ \eproof

Now, in running the regressions, we do not make use of the details of the formula in the proof. We merely note that there is an affine bias if there is some homophily inside the islands on dimensions falling outside the scope of our data. Thus, including an intercept in the regression of convergence rates on $\tilde{x}$ is a reasonable first-order approximation to account for some of this affine bias.

Of course, in more realistic settings, the various kinds of symmetry present in this model will not exist. However, it appears that more general formulas or characterizations could be obtained describing how homophilies at various levels interact. This could be a useful direction to pursue in taking this model to empirical settings, where there will almost always be some underlying homophily on dimensions not captured by the data.

\subsubsection{The Asymptotic Equivalence of DWH and EDWH}

\noindent{\bf Proof of Lemma \ref{lem:dwhedwh}}:
Consider the islands model with $m(n)\geq 2$ equal-sized groups, and probabilities of links within and across types
$p_s$ and $p_d$, respectively, and 
consider any sequence of groupings of islands ${M(n)\in I(n)}$.
We show that
$$|\EDWH(M(n),m,p_s,p_d) -   \DWH(M(n),\mathbf{A(P,n)}) |\xrightarrow{p} 0.$$
This follows from showing that $|EW_{M(n),M(n)} - W_{M(n)M(n)}(\mathbf{A(P,n)})|\xrightarrow{p} 0,$  $|EW_{M(n),M^c(n)} - W_{M(n)M^c(n)}(\mathbf{A(P,n)})|\xrightarrow{p} 0,$
and 
$(\sum_{i \in M(n)} \frac{1}{d(n)}) /(\frac{1}{d_i(\mathbf{A(P,n)})} )\xrightarrow{p} 1,$ for any $M(n)$ (and noting that the denominator is bounded away from 0 in the limit).

The latter conclusion follows from the argument in the proof of
Corollary \ref{cor:islandslimits} using Lemma 2.1 in Chung and Lu (2002) to show that with a probability going to 1 all nodes have degree between
$(1-f(n))d(n)$ and $(1+f(n))d(n)$ for any function $f(n)\rightarrow 0$ such that $(f(n))^2\log(n)/ 3$ goes to $\infty$.  
Next, given that $m(n)/n\rightarrow 0$ which implies that the number of nodes in within any island $i(n)=n/m(n)$ is growing without bound, and so we can again apply Lemma 2.1 in Chung and Lu (2002),\footnote{Now,
we work with the $X_i$'s in their lemma to be the realization of the links within a given $M(n)$.}
to deduce that there is a function $g(n)\rightarrow 0$ such that with a probability going to 1, 
$$(1-g(n))\left[k(n)p_s(n) |i(n)|^2/d(n)^2 + p_d(n) k(n)(k(n)-1) |i(n)|^2/d(n)^2\right ] \leq \sum_{i\in M, j\in M} T_{ij}T_{ji}  $$
$$ \leq (1+g(n))\left[k(n)p_s(n) |i(n)|^2/d(n)^2 + p_d(n) k(n)(k(n)-1) |i(n)|^2/d(n)^2\right ] ,$$ 
where $k(n)$ is the number of islands in $M(n)$
and similarly 
$$(1-g(n))p_d(n) |M(n)| |M^c(n)|/d(n)^2 \leq \sum_{i\in M, j\in M} T_{ij}T_{ji}  \leq (1+g(n))p_d(n) |M(n)| |M^c(n)|/d(n)^2.$$ 
These imply that $$|EW_{M(n),M(n)} - W_{M(n)M(n)}(\mathbf{A(P,n)})|\xrightarrow{p} 0$$ and  $$|EW_{M(n),M^c(n)} - W_{M(n)M^c(n)}(\mathbf{A(P,n)})|\xrightarrow{p} 0,$$ as claimed.\eproof

\subsection{Results of the Empirical Analysis for Mixing Time}

{
\def\sep{0.5em}
\def\fns{\footnotesize}
\def\onepc{$^{\ast\ast}$} \def\fivepc{$^{\ast}$}
\def\tenpc{$^{\dag}$}
\def\legend{\multicolumn{3}{l}{\footnotesize{Significance levels
:\hspace{1em} $\dag$ : 10\% \hspace{1em}
$\ast$ : 5\% \hspace{1em} $\ast\ast$ : 1\% \normalsize}}}
\begin{table}[htbp]\centering
 \caption{Dependent variable =  $\rho(\MT(0.1/n;\mathbf{A}))$} ($N=82$)\\
\label{tab:rhoMTonHtype}
\begin{tabular}{l r l}\hline\hline 
\textbf{Variable}  & \multicolumn{1}{c}{\textbf{Coefficient}} &\\
  & \multicolumn{1}{c}{\textbf{($t$-statistic)}} &\\\hline
Intercept & 0.861 \\ &\fns{(66.9)}  \\[\sep]
$\frac{H-1}{m-1}$ for ``type'' & 0.287  \\ &\fns{(5.23)} \\[\sep]
\hline R$^{2}$ & {0.255}  &\\
\hline
\hline
\end{tabular}
\end{table}
}

{
\def\sep{0.5em}
\def\fns{\footnotesize}
\def\onepc{$^{\ast\ast}$} \def\fivepc{$^{\ast}$}
\def\tenpc{$^{\dag}$}
\def\legend{\multicolumn{3}{l}{\footnotesize{Significance levels
:\hspace{1em} $\dag$ : 10\% \hspace{1em}
$\ast$ : 5\% \hspace{1em} $\ast\ast$ : 1\% \normalsize}}}
\begin{table}[htbp]\centering
 \caption{Dependent variable =  $\rho(\MT(0.1/n;\mathbf{A}))$} ($N=82$)\\
\label{tab:rhoMTonH}
\begin{tabular}{l r l}\hline\hline 
\textbf{Variable}  & \multicolumn{1}{c}{\textbf{Coefficient}} &\\
  & \multicolumn{1}{c}{\textbf{($t$-statistic)}} &\\\hline
Intercept & 0.825 \\ &\fns{(34.6)}  \\[\sep]
$\frac{H-1}{m-1}$ for grade & 0.163  \\ &\fns{(4.24)} \\[\sep]
\hline R$^{2}$ & {0.184}  &\\
\hline
\hline
\end{tabular}
\end{table}
}

{
\def\sep{0.5em}
\def\fns{\footnotesize}
\def\onepc{$^{\ast\ast}$} \def\fivepc{$^{\ast}$}
\def\tenpc{$^{\dag}$}
\def\legend{\multicolumn{3}{l}{\footnotesize{Significance levels
:\hspace{1em} $\dag$ : 10\% \hspace{1em}
$\ast$ : 5\% \hspace{1em} $\ast\ast$ : 1\% \normalsize}}}
\begin{table}[htbp]\centering 
\caption{Dependent variable = Imputed convergence rate, $\rho(\MT(0.1/n;\mathbf{A}))$} ($N=82$) \\
\label{tab:rhoMT}
\begin{tabular}{l r r r r l}\hline\hline 
\textbf{Variable}  & \multicolumn{4}{c}{\textbf{Coefficient}} &\\
  & \multicolumn{4}{c}{\textbf{($t$-statistic)}} &\\\hline
&  \emph{All Homophilies}  & \emph{Grade Only} & \emph{Gender Only} & \emph{Race Only} &\\[\sep]
Intercept & 0.665 & 0.735 & 0.884 & 0.902 \\ &\fns{(22.4)} & \fns{(23.3)} &\fns{(60.2)} & \fns{(95.6)} \\[\sep]
Grade DWH & 0.309 & 0.284 & --- & ---  \\ &\fns{(7.26)} & \fns{(6.05)} &\fns{} & \fns{}\\[\sep]
Gender DWH & 0.100 & ---  & 0.183 & ---\\ & \fns{(2.02)} & \fns{} & \fns{(2.84)} & \fns{} \\[\sep]
Race DWH & 0.104 & --- & --- & 0.0696 \\ & \fns{(5.23)} & \fns{} & \fns{} & \fns{(2.65)} \\[\sep]
\hline R$^{2}$ & {0.511} & {0.314}  & {0.0917} & {0.0807} &\\
\hline
\hline
\end{tabular}
\end{table}
}

\begin{figure}[htp] \begin{center}
\includegraphics[width=5in]{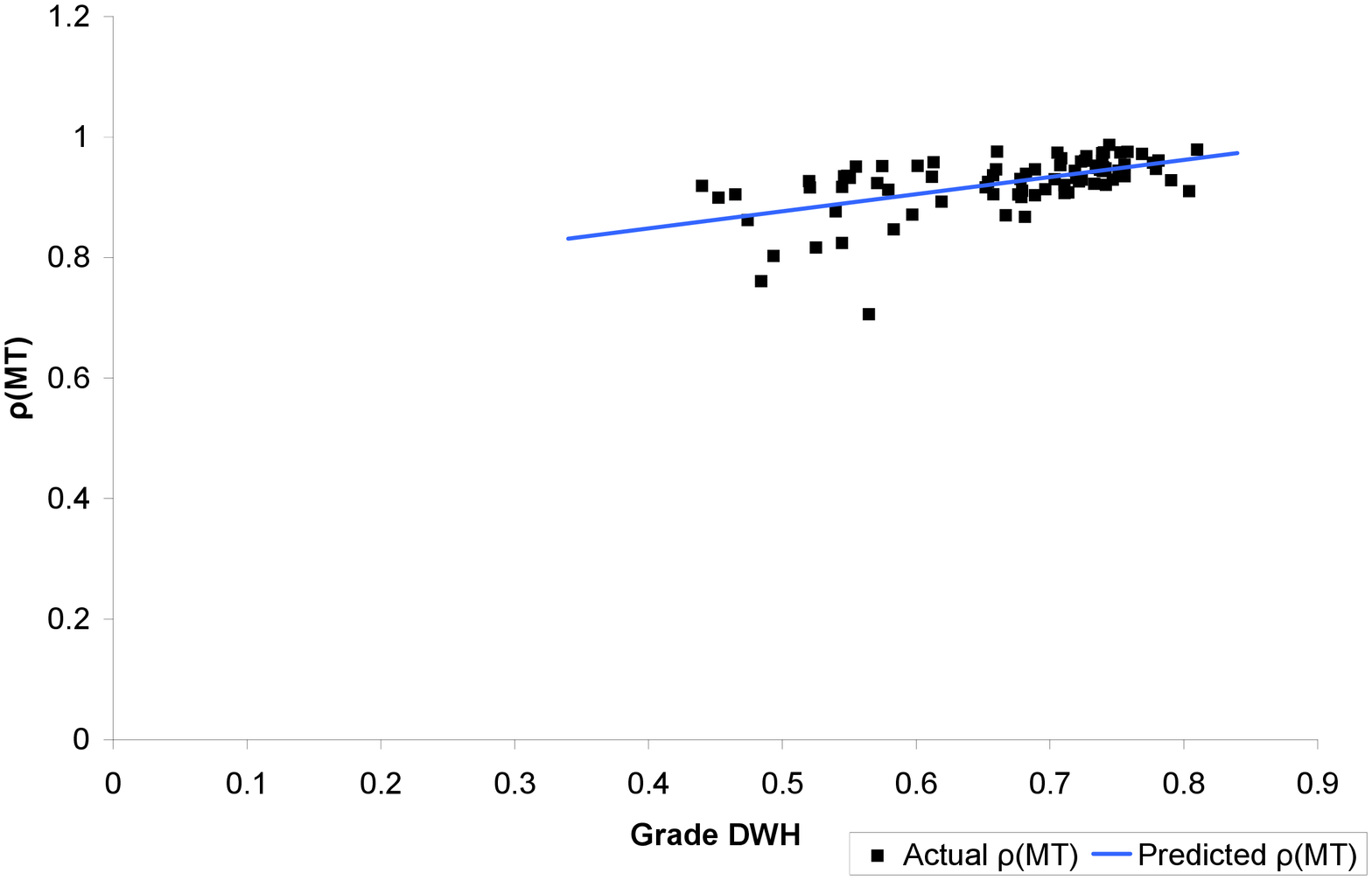}
%\begin{enumerate}
%	\item 
%\end{enumerate}

\caption{Rate of convergence of mixing time for the 82 friendship networks plotted against the degree weighted homophily in each of the networks calculated
relative to grade.} 
\label{fig:rhoMTvsGradeDWH}
\end{center}
\end{figure}

\end{document}